
\catcode`\@=11


\message{Loading jyTeX fonts...}



\font\vptrm=cmr5 \font\vptmit=cmmi5 \font\vptsy=cmsy5 \font\vptbf=cmbx5

\skewchar\vptmit='177 \skewchar\vptsy='60 \fontdimen16
\vptsy=\the\fontdimen17 \vptsy

\def\vpt{\ifmmode\err@badsizechange\else
     \@mathfontinit
     \textfont0=\vptrm  \scriptfont0=\vptrm  \scriptscriptfont0=\vptrm
     \textfont1=\vptmit \scriptfont1=\vptmit \scriptscriptfont1=\vptmit
     \textfont2=\vptsy  \scriptfont2=\vptsy  \scriptscriptfont2=\vptsy
     \textfont3=\xptex  \scriptfont3=\xptex  \scriptscriptfont3=\xptex
     \textfont\bffam=\vptbf
     \scriptfont\bffam=\vptbf
     \scriptscriptfont\bffam=\vptbf
     \@fontstyleinit
     \def\rm{\vptrm\fam=\z@}%
     \def\bf{\vptbf\fam=\bffam}%
     \def\oldstyle{\vptmit\fam=\@ne}%
     \rm\fi}


\font\viptrm=cmr6 \font\viptmit=cmmi6 \font\viptsy=cmsy6
\font\viptbf=cmbx6

\skewchar\viptmit='177 \skewchar\viptsy='60 \fontdimen16
\viptsy=\the\fontdimen17 \viptsy

\def\vipt{\ifmmode\err@badsizechange\else
     \@mathfontinit
     \textfont0=\viptrm  \scriptfont0=\vptrm  \scriptscriptfont0=\vptrm
     \textfont1=\viptmit \scriptfont1=\vptmit \scriptscriptfont1=\vptmit
     \textfont2=\viptsy  \scriptfont2=\vptsy  \scriptscriptfont2=\vptsy
     \textfont3=\xptex   \scriptfont3=\xptex  \scriptscriptfont3=\xptex
     \textfont\bffam=\viptbf
     \scriptfont\bffam=\vptbf
     \scriptscriptfont\bffam=\vptbf
     \@fontstyleinit
     \def\rm{\viptrm\fam=\z@}%
     \def\bf{\viptbf\fam=\bffam}%
     \def\oldstyle{\viptmit\fam=\@ne}%
     \rm\fi}

\font\viiptrm=cmr7 \font\viiptmit=cmmi7 \font\viiptsy=cmsy7
\font\viiptit=cmti7 \font\viiptbf=cmbx7

\skewchar\viiptmit='177 \skewchar\viiptsy='60 \fontdimen16
\viiptsy=\the\fontdimen17 \viiptsy

\def\viipt{\ifmmode\err@badsizechange\else
     \@mathfontinit
     \textfont0=\viiptrm  \scriptfont0=\vptrm  \scriptscriptfont0=\vptrm
     \textfont1=\viiptmit \scriptfont1=\vptmit \scriptscriptfont1=\vptmit
     \textfont2=\viiptsy  \scriptfont2=\vptsy  \scriptscriptfont2=\vptsy
     \textfont3=\xptex    \scriptfont3=\xptex  \scriptscriptfont3=\xptex
     \textfont\itfam=\viiptit
     \scriptfont\itfam=\viiptit
     \scriptscriptfont\itfam=\viiptit
     \textfont\bffam=\viiptbf
     \scriptfont\bffam=\vptbf
     \scriptscriptfont\bffam=\vptbf
     \@fontstyleinit
     \def\rm{\viiptrm\fam=\z@}%
     \def\it{\viiptit\fam=\itfam}%
     \def\bf{\viiptbf\fam=\bffam}%
     \def\oldstyle{\viiptmit\fam=\@ne}%
     \rm\fi}


\font\viiiptrm=cmr8 \font\viiiptmit=cmmi8 \font\viiiptsy=cmsy8
\font\viiiptit=cmti8
\font\viiiptbf=cmbx8

\skewchar\viiiptmit='177 \skewchar\viiiptsy='60 \fontdimen16
\viiiptsy=\the\fontdimen17 \viiiptsy

\def\viiipt{\ifmmode\err@badsizechange\else
     \@mathfontinit
     \textfont0=\viiiptrm  \scriptfont0=\viptrm  \scriptscriptfont0=\vptrm
     \textfont1=\viiiptmit \scriptfont1=\viptmit \scriptscriptfont1=\vptmit
     \textfont2=\viiiptsy  \scriptfont2=\viptsy  \scriptscriptfont2=\vptsy
     \textfont3=\xptex     \scriptfont3=\xptex   \scriptscriptfont3=\xptex
     \textfont\itfam=\viiiptit
     \scriptfont\itfam=\viiptit
     \scriptscriptfont\itfam=\viiptit
     \textfont\bffam=\viiiptbf
     \scriptfont\bffam=\viptbf
     \scriptscriptfont\bffam=\vptbf
     \@fontstyleinit
     \def\rm{\viiiptrm\fam=\z@}%
     \def\it{\viiiptit\fam=\itfam}%
     \def\bf{\viiiptbf\fam=\bffam}%
     \def\oldstyle{\viiiptmit\fam=\@ne}%
     \rm\fi}


\def\getixpt{%
     \font\ixptrm=cmr9
     \font\ixptmit=cmmi9
     \font\ixptsy=cmsy9
     \font\ixptit=cmti9
     \font\ixptbf=cmbx9
     \skewchar\ixptmit='177 \skewchar\ixptsy='60
     \fontdimen16 \ixptsy=\the\fontdimen17 \ixptsy}

\def\ixpt{\ifmmode\err@badsizechange\else
     \@mathfontinit
     \textfont0=\ixptrm  \scriptfont0=\viiptrm  \scriptscriptfont0=\vptrm
     \textfont1=\ixptmit \scriptfont1=\viiptmit \scriptscriptfont1=\vptmit
     \textfont2=\ixptsy  \scriptfont2=\viiptsy  \scriptscriptfont2=\vptsy
     \textfont3=\xptex   \scriptfont3=\xptex    \scriptscriptfont3=\xptex
     \textfont\itfam=\ixptit
     \scriptfont\itfam=\viiptit
     \scriptscriptfont\itfam=\viiptit
     \textfont\bffam=\ixptbf
     \scriptfont\bffam=\viiptbf
     \scriptscriptfont\bffam=\vptbf
     \@fontstyleinit
     \def\rm{\ixptrm\fam=\z@}%
     \def\it{\ixptit\fam=\itfam}%
     \def\bf{\ixptbf\fam=\bffam}%
     \def\oldstyle{\ixptmit\fam=\@ne}%
     \rm\fi}


\font\xptrm=cmr10 \font\xptmit=cmmi10 \font\xptsy=cmsy10
\font\xptex=cmex10 \font\xptit=cmti10 \font\xptsl=cmsl10
\font\xptbf=cmbx10 \font\xpttt=cmtt10 \font\xptss=cmss10
\font\xptsc=cmcsc10 \font\xptbfs=cmb10 \font\xptbmit=cmmib10

\skewchar\xptmit='177 \skewchar\xptbmit='177 \skewchar\xptsy='60
\fontdimen16 \xptsy=\the\fontdimen17 \xptsy

\def\xpt{\ifmmode\err@badsizechange\else
     \@mathfontinit
     \textfont0=\xptrm  \scriptfont0=\viiptrm  \scriptscriptfont0=\vptrm
     \textfont1=\xptmit \scriptfont1=\viiptmit \scriptscriptfont1=\vptmit
     \textfont2=\xptsy  \scriptfont2=\viiptsy  \scriptscriptfont2=\vptsy
     \textfont3=\xptex  \scriptfont3=\xptex    \scriptscriptfont3=\xptex
     \textfont\itfam=\xptit
     \scriptfont\itfam=\viiptit
     \scriptscriptfont\itfam=\viiptit
     \textfont\bffam=\xptbf
     \scriptfont\bffam=\viiptbf
     \scriptscriptfont\bffam=\vptbf
     \textfont\bfsfam=\xptbfs
     \scriptfont\bfsfam=\viiptbf
     \scriptscriptfont\bfsfam=\vptbf
     \textfont\bmitfam=\xptbmit
     \scriptfont\bmitfam=\viiptmit
     \scriptscriptfont\bmitfam=\vptmit
     \@fontstyleinit
     \def\rm{\xptrm\fam=\z@}%
     \def\it{\xptit\fam=\itfam}%
     \def\sl{\xptsl}%
     \def\bf{\xptbf\fam=\bffam}%
     \def\tt{\xpttt}%
     \def\ss{\xptss}%
     \def\sc{\xptsc}%
     \def\bfs{\xptbfs\fam=\bfsfam}%
     \def\bmit{\fam=\bmitfam}%
     \def\oldstyle{\xptmit\fam=\@ne}%
     \rm\fi}


\def\getxipt{%
     \font\xiptrm=cmr10  scaled\magstephalf
     \font\xiptmit=cmmi10 scaled\magstephalf
     \font\xiptsy=cmsy10 scaled\magstephalf
     \font\xiptex=cmex10 scaled\magstephalf
     \font\xiptit=cmti10 scaled\magstephalf
     \font\xiptsl=cmsl10 scaled\magstephalf
     \font\xiptbf=cmbx10 scaled\magstephalf
     \font\xipttt=cmtt10 scaled\magstephalf
     \font\xiptss=cmss10 scaled\magstephalf
     \skewchar\xiptmit='177 \skewchar\xiptsy='60
     \fontdimen16 \xiptsy=\the\fontdimen17 \xiptsy}

\def\xipt{\ifmmode\err@badsizechange\else
     \@mathfontinit
     \textfont0=\xiptrm  \scriptfont0=\viiiptrm  \scriptscriptfont0=\viptrm
     \textfont1=\xiptmit \scriptfont1=\viiiptmit \scriptscriptfont1=\viptmit
     \textfont2=\xiptsy  \scriptfont2=\viiiptsy  \scriptscriptfont2=\viptsy
     \textfont3=\xiptex  \scriptfont3=\xptex     \scriptscriptfont3=\xptex
     \textfont\itfam=\xiptit
     \scriptfont\itfam=\viiiptit
     \scriptscriptfont\itfam=\viiptit
     \textfont\bffam=\xiptbf
     \scriptfont\bffam=\viiiptbf
     \scriptscriptfont\bffam=\viptbf
     \@fontstyleinit
     \def\rm{\xiptrm\fam=\z@}%
     \def\it{\xiptit\fam=\itfam}%
     \def\sl{\xiptsl}%
     \def\bf{\xiptbf\fam=\bffam}%
     \def\tt{\xipttt}%
     \def\ss{\xiptss}%
     \def\oldstyle{\xiptmit\fam=\@ne}%
     \rm\fi}


\font\xiiptrm=cmr12 \font\xiiptmit=cmmi12 \font\xiiptsy=cmsy10
scaled\magstep1 \font\xiiptex=cmex10  scaled\magstep1
\font\xiiptit=cmti12 \font\xiiptsl=cmsl12 \font\xiiptbf=cmbx12
\font\xiiptss=cmss12 \font\xiiptsc=cmcsc10 scaled\magstep1
\font\xiiptbfs=cmb10  scaled\magstep1 \font\xiiptbmit=cmmib10
scaled\magstep1

\skewchar\xiiptmit='177 \skewchar\xiiptbmit='177 \skewchar\xiiptsy='60
\fontdimen16 \xiiptsy=\the\fontdimen17 \xiiptsy

\def\xiipt{\ifmmode\err@badsizechange\else
     \@mathfontinit
     \textfont0=\xiiptrm  \scriptfont0=\viiiptrm  \scriptscriptfont0=\viptrm
     \textfont1=\xiiptmit \scriptfont1=\viiiptmit \scriptscriptfont1=\viptmit
     \textfont2=\xiiptsy  \scriptfont2=\viiiptsy  \scriptscriptfont2=\viptsy
     \textfont3=\xiiptex  \scriptfont3=\xptex     \scriptscriptfont3=\xptex
     \textfont\itfam=\xiiptit
     \scriptfont\itfam=\viiiptit
     \scriptscriptfont\itfam=\viiptit
     \textfont\bffam=\xiiptbf
     \scriptfont\bffam=\viiiptbf
     \scriptscriptfont\bffam=\viptbf
     \textfont\bfsfam=\xiiptbfs
     \scriptfont\bfsfam=\viiiptbf
     \scriptscriptfont\bfsfam=\viptbf
     \textfont\bmitfam=\xiiptbmit
     \scriptfont\bmitfam=\viiiptmit
     \scriptscriptfont\bmitfam=\viptmit
     \@fontstyleinit
     \def\rm{\xiiptrm\fam=\z@}%
     \def\it{\xiiptit\fam=\itfam}%
     \def\sl{\xiiptsl}%
     \def\bf{\xiiptbf\fam=\bffam}%
     \def\tt{\xiipttt}%
     \def\ss{\xiiptss}%
     \def\sc{\xiiptsc}%
     \def\bfs{\xiiptbfs\fam=\bfsfam}%
     \def\bmit{\fam=\bmitfam}%
     \def\oldstyle{\xiiptmit\fam=\@ne}%
     \rm\fi}


\def\getxiiipt{%
     \font\xiiiptrm=cmr12  scaled\magstephalf
     \font\xiiiptmit=cmmi12 scaled\magstephalf
     \font\xiiiptsy=cmsy9  scaled\magstep2
     \font\xiiiptit=cmti12 scaled\magstephalf
     \font\xiiiptsl=cmsl12 scaled\magstephalf
     \font\xiiiptbf=cmbx12 scaled\magstephalf
     \font\xiiipttt=cmtt12 scaled\magstephalf
     \font\xiiiptss=cmss12 scaled\magstephalf
     \skewchar\xiiiptmit='177 \skewchar\xiiiptsy='60
     \fontdimen16 \xiiiptsy=\the\fontdimen17 \xiiiptsy}

\def\xiiipt{\ifmmode\err@badsizechange\else
     \@mathfontinit
     \textfont0=\xiiiptrm  \scriptfont0=\xptrm  \scriptscriptfont0=\viiptrm
     \textfont1=\xiiiptmit \scriptfont1=\xptmit \scriptscriptfont1=\viiptmit
     \textfont2=\xiiiptsy  \scriptfont2=\xptsy  \scriptscriptfont2=\viiptsy
     \textfont3=\xivptex   \scriptfont3=\xptex  \scriptscriptfont3=\xptex
     \textfont\itfam=\xiiiptit
     \scriptfont\itfam=\xptit
     \scriptscriptfont\itfam=\viiptit
     \textfont\bffam=\xiiiptbf
     \scriptfont\bffam=\xptbf
     \scriptscriptfont\bffam=\viiptbf
     \@fontstyleinit
     \def\rm{\xiiiptrm\fam=\z@}%
     \def\it{\xiiiptit\fam=\itfam}%
     \def\sl{\xiiiptsl}%
     \def\bf{\xiiiptbf\fam=\bffam}%
     \def\tt{\xiiipttt}%
     \def\ss{\xiiiptss}%
     \def\oldstyle{\xiiiptmit\fam=\@ne}%
     \rm\fi}


\font\xivptrm=cmr12   scaled\magstep1 \font\xivptmit=cmmi12
scaled\magstep1 \font\xivptsy=cmsy10  scaled\magstep2
\font\xivptex=cmex10  scaled\magstep2 \font\xivptit=cmti12
scaled\magstep1 \font\xivptsl=cmsl12  scaled\magstep1
\font\xivptbf=cmbx12  scaled\magstep1
\font\xivptss=cmss12  scaled\magstep1 \font\xivptsc=cmcsc10
scaled\magstep2 \font\xivptbfs=cmb10  scaled\magstep2
\font\xivptbmit=cmmib10 scaled\magstep2

\skewchar\xivptmit='177 \skewchar\xivptbmit='177 \skewchar\xivptsy='60
\fontdimen16 \xivptsy=\the\fontdimen17 \xivptsy

\def\xivpt{\ifmmode\err@badsizechange\else
     \@mathfontinit
     \textfont0=\xivptrm  \scriptfont0=\xptrm  \scriptscriptfont0=\viiptrm
     \textfont1=\xivptmit \scriptfont1=\xptmit \scriptscriptfont1=\viiptmit
     \textfont2=\xivptsy  \scriptfont2=\xptsy  \scriptscriptfont2=\viiptsy
     \textfont3=\xivptex  \scriptfont3=\xptex  \scriptscriptfont3=\xptex
     \textfont\itfam=\xivptit
     \scriptfont\itfam=\xptit
     \scriptscriptfont\itfam=\viiptit
     \textfont\bffam=\xivptbf
     \scriptfont\bffam=\xptbf
     \scriptscriptfont\bffam=\viiptbf
     \textfont\bfsfam=\xivptbfs
     \scriptfont\bfsfam=\xptbfs
     \scriptscriptfont\bfsfam=\viiptbf
     \textfont\bmitfam=\xivptbmit
     \scriptfont\bmitfam=\xptbmit
     \scriptscriptfont\bmitfam=\viiptmit
     \@fontstyleinit
     \def\rm{\xivptrm\fam=\z@}%
     \def\it{\xivptit\fam=\itfam}%
     \def\sl{\xivptsl}%
     \def\bf{\xivptbf\fam=\bffam}%
     \def\tt{\xivpttt}%
     \def\ss{\xivptss}%
     \def\sc{\xivptsc}%
     \def\bfs{\xivptbfs\fam=\bfsfam}%
     \def\bmit{\fam=\bmitfam}%
     \def\oldstyle{\xivptmit\fam=\@ne}%
     \rm\fi}


\font\xviiptrm=cmr17 \font\xviiptmit=cmmi12 scaled\magstep2
\font\xviiptsy=cmsy10 scaled\magstep3 \font\xviiptex=cmex10
scaled\magstep3 \font\xviiptit=cmti12 scaled\magstep2
\font\xviiptbf=cmbx12 scaled\magstep2 \font\xviiptbfs=cmb10
scaled\magstep3

\skewchar\xviiptmit='177 \skewchar\xviiptsy='60 \fontdimen16
\xviiptsy=\the\fontdimen17 \xviiptsy

\def\xviipt{\ifmmode\err@badsizechange\else
     \@mathfontinit
     \textfont0=\xviiptrm  \scriptfont0=\xiiptrm  \scriptscriptfont0=\viiiptrm
     \textfont1=\xviiptmit \scriptfont1=\xiiptmit \scriptscriptfont1=\viiiptmit
     \textfont2=\xviiptsy  \scriptfont2=\xiiptsy  \scriptscriptfont2=\viiiptsy
     \textfont3=\xviiptex  \scriptfont3=\xiiptex  \scriptscriptfont3=\xptex
     \textfont\itfam=\xviiptit
     \scriptfont\itfam=\xiiptit
     \scriptscriptfont\itfam=\viiiptit
     \textfont\bffam=\xviiptbf
     \scriptfont\bffam=\xiiptbf
     \scriptscriptfont\bffam=\viiiptbf
     \textfont\bfsfam=\xviiptbfs
     \scriptfont\bfsfam=\xiiptbfs
     \scriptscriptfont\bfsfam=\viiiptbf
     \@fontstyleinit
     \def\rm{\xviiptrm\fam=\z@}%
     \def\it{\xviiptit\fam=\itfam}%
     \def\bf{\xviiptbf\fam=\bffam}%
     \def\bfs{\xviiptbfs\fam=\bfsfam}%
     \def\oldstyle{\xviiptmit\fam=\@ne}%
     \rm\fi}


\font\xxiptrm=cmr17  scaled\magstep1


\def\xxipt{\ifmmode\err@badsizechange\else
     \@mathfontinit
     \@fontstyleinit
     \def\rm{\xxiptrm\fam=\z@}%
     \rm\fi}


\font\xxvptrm=cmr17  scaled\magstep2


\def\xxvpt{\ifmmode\err@badsizechange\else
     \@mathfontinit
     \@fontstyleinit
     \def\rm{\xxvptrm\fam=\z@}%
     \rm\fi}




\message{Loading jyTeX macros...}

\message{modifications to plain.tex,}


\def\newcount{\alloc@0\count\countdef\insc@unt}
\def\newdimen{\alloc@1\dimen\dimendef\insc@unt}
\def\newskip{\alloc@2\skip\skipdef\insc@unt}
\def\newmuskip{\alloc@3\muskip\muskipdef\@cclvi}
\def\newbox{\alloc@4\box\chardef\insc@unt}
\def\newtoks{\alloc@5\toks\toksdef\@cclvi}
\def\newhelp#1#2{\newtoks#1\global#1\expandafter{\csname#2\endcsname}}
\def\newread{\alloc@6\read\chardef\sixt@@n}
\def\newwrite{\alloc@7\write\chardef\sixt@@n}
\def\newfam{\alloc@8\fam\chardef\sixt@@n}
\def\newinsert#1{\global\advance\insc@unt by\m@ne
     \ch@ck0\insc@unt\count
     \ch@ck1\insc@unt\dimen
     \ch@ck2\insc@unt\skip
     \ch@ck4\insc@unt\box
     \allocationnumber=\insc@unt
     \global\chardef#1=\allocationnumber
     \wlog{\string#1=\string\insert\the\allocationnumber}}
\def\newif#1{\count@\escapechar \escapechar\m@ne
     \expandafter\expandafter\expandafter
          \xdef\@if#1{true}{\let\noexpand#1=\noexpand\iftrue}%
     \expandafter\expandafter\expandafter
          \xdef\@if#1{false}{\let\noexpand#1=\noexpand\iffalse}%
     \global\@if#1{false}\escapechar=\count@}


\newlinechar=`\^^J
\overfullrule=0pt




\let\itfam=\undefined

\let\bffam=\undefined

\count18=3


\chardef\sharps="19


\mathchardef\alpha="710B \mathchardef\beta="710C \mathchardef\gamma="710D
\mathchardef\delta="710E \mathchardef\epsilon="710F
\mathchardef\zeta="7110 \mathchardef\eta="7111 \mathchardef\theta="7112
\mathchardef\iota="7113 \mathchardef\kappa="7114
\mathchardef\lambda="7115 \mathchardef\mu="7116 \mathchardef\nu="7117
\mathchardef\xi="7118 \mathchardef\pi="7119 \mathchardef\rho="711A
\mathchardef\sigma="711B \mathchardef\tau="711C
\mathchardef\upsilon="711D \mathchardef\phi="711E \mathchardef\chi="711F
\mathchardef\psi="7120 \mathchardef\omega="7121
\mathchardef\varepsilon="7122 \mathchardef\vartheta="7123
\mathchardef\varpi="7124 \mathchardef\varrho="7125
\mathchardef\varsigma="7126 \mathchardef\varphi="7127
\mathchardef\imath="717B \mathchardef\jmath="717C \mathchardef\ell="7160
\mathchardef\wp="717D \mathchardef\partial="7140 \mathchardef\flat="715B
\mathchardef\natural="715C \mathchardef\sharp="715D



\def\angle{{\vbox{\ialign{$\m@th\scriptstyle##$\crcr
     \not\mathrel{\mkern14mu}\crcr
     \noalign{\nointerlineskip}
     \mkern2.5mu\leaders\hrule height.34\rp@\hfill\mkern2.5mu\crcr}}}}
\def\vdots{\vbox{\baselineskip4\rp@ \lineskiplimit\z@
     \kern6\rp@\hbox{.}\hbox{.}\hbox{.}}}
\def\ddots{\mathinner{\mkern1mu\raise7\rp@\vbox{\kern7\rp@\hbox{.}}\mkern2mu
     \raise4\rp@\hbox{.}\mkern2mu\raise\rp@\hbox{.}\mkern1mu}}
\def\overrightarrow#1{\vbox{\ialign{##\crcr
     \rightarrowfill\crcr
     \noalign{\kern-\rp@\nointerlineskip}
     $\hfil\displaystyle{#1}\hfil$\crcr}}}
\def\overleftarrow#1{\vbox{\ialign{##\crcr
     \leftarrowfill\crcr
     \noalign{\kern-\rp@\nointerlineskip}
     $\hfil\displaystyle{#1}\hfil$\crcr}}}
\def\overbrace#1{\mathop{\vbox{\ialign{##\crcr
     \noalign{\kern3\rp@}
     \downbracefill\crcr
     \noalign{\kern3\rp@\nointerlineskip}
     $\hfil\displaystyle{#1}\hfil$\crcr}}}\limits}
\def\underbrace#1{\mathop{\vtop{\ialign{##\crcr
     $\hfil\displaystyle{#1}\hfil$\crcr
     \noalign{\kern3\rp@\nointerlineskip}
     \upbracefill\crcr
     \noalign{\kern3\rp@}}}}\limits}
\def\big#1{{\hbox{$\left#1\vbox to8.5\rp@ {}\right.\n@space$}}}
\def\Big#1{{\hbox{$\left#1\vbox to11.5\rp@ {}\right.\n@space$}}}
\def\bigg#1{{\hbox{$\left#1\vbox to14.5\rp@ {}\right.\n@space$}}}
\def\Bigg#1{{\hbox{$\left#1\vbox to17.5\rp@ {}\right.\n@space$}}}
\def\@vereq#1#2{\lower.5\rp@\vbox{\baselineskip\z@skip\lineskip-.5\rp@
     \ialign{$\m@th#1\hfil##\hfil$\crcr#2\crcr=\crcr}}}
\def\rlh@#1{\vcenter{\hbox{\ooalign{\raise2\rp@
     \hbox{$#1\rightharpoonup$}\crcr
     $#1\leftharpoondown$}}}}
\def\bordermatrix#1{\begingroup\m@th
     \setbox\z@\vbox{%
          \def\cr{\crcr\noalign{\kern2\rp@\global\let\cr\endline}}%
          \ialign{$##$\hfil\kern2\rp@\kern\p@renwd
               &\thinspace\hfil$##$\hfil&&\quad\hfil$##$\hfil\crcr
               \omit\strut\hfil\crcr
               \noalign{\kern-\baselineskip}%
               #1\crcr\omit\strut\cr}}%
     \setbox\tw@\vbox{\unvcopy\z@\global\setbox\@ne\lastbox}%
     \setbox\tw@\hbox{\unhbox\@ne\unskip\global\setbox\@ne\lastbox}%
     \setbox\tw@\hbox{$\kern\wd\@ne\kern-\p@renwd\left(\kern-\wd\@ne
          \global\setbox\@ne\vbox{\box\@ne\kern2\rp@}%
          \vcenter{\kern-\ht\@ne\unvbox\z@\kern-\baselineskip}%
          \,\right)$}%
     \null\;\vbox{\kern\ht\@ne\box\tw@}\endgroup}
\def\endinsert{\egroup
     \if@mid\dimen@\ht\z@
          \advance\dimen@\dp\z@
          \advance\dimen@12\rp@
          \advance\dimen@\pagetotal
          \ifdim\dimen@>\pagegoal\@midfalse\p@gefalse\fi
     \fi
     \if@mid\bigskip\box\z@
          \bigbreak
     \else\insert\topins{\penalty100 \splittopskip\z@skip
               \splitmaxdepth\maxdimen\floatingpenalty\z@
               \ifp@ge\dimen@\dp\z@
                    \vbox to\vsize{\unvbox\z@\kern-\dimen@}%
               \else\box\z@\nobreak\bigskip
               \fi}%
     \fi
     \endgroup}


\def\cases#1{\left\{\,\vcenter{\m@th
     \ialign{$##\hfil$&\quad##\hfil\crcr#1\crcr}}\right.}
\def\matrix#1{\null\,\vcenter{\m@th
     \ialign{\hfil$##$\hfil&&\quad\hfil$##$\hfil\crcr
          \mathstrut\crcr
          \noalign{\kern-\baselineskip}
          #1\crcr
          \mathstrut\crcr
          \noalign{\kern-\baselineskip}}}\,}


\newif\ifraggedbottom

\def\raggedbottom{\ifraggedbottom\else
     \advance\topskip by\z@ plus60pt \raggedbottomtrue\fi}%
\def\normalbottom{\ifraggedbottom
     \advance\topskip by\z@ plus-60pt \raggedbottomfalse\fi}

\message{hacks,}


\toksdef\toks@i=1 \toksdef\toks@ii=2


\def\TeX{T\kern-.1667em \lower.5ex \hbox{E}\kern-.125em X\null}
\def\jyTeX{{\leavevmode
     \raise.587ex \hbox{\it\j}\kern-.1em \lower.048ex \hbox{\it y}\kern-.12em
     \TeX}}

\let\then=\iftrue
\def\ifnoarg#1\then{\def\hack@{#1}\ifx\hack@\empty}
\def\ifundefined#1\then{%
     \expandafter\ifx\csname\expandafter\blank\string#1\endcsname\relax}
\def\useif#1\then{\csname#1\endcsname}
\def\usename#1{\csname#1\endcsname}
\def\useafter#1#2{\expandafter#1\csname#2\endcsname}

\long\def\loop#1\repeat{\def\@iterate{#1\expandafter\@iterate\fi}\@iterate
     \let\@iterate=\relax}

\let\TeXend=\end
\def\begin#1{\begingroup\def\@@blockname{#1}\usename{begin#1}}
\def\end#1{\usename{end#1}\def\hack@{#1}%
     \ifx\@@blockname\hack@
          \endgroup
     \else\err@badgroup\hack@\@@blockname
     \fi}
\def\@@blockname{}

\def\defaultoption[#1]#2{%
     \def\hack@{\ifx\hack@ii[\toks@={#2}\else\toks@={#2[#1]}\fi\the\toks@}%
     \futurelet\hack@ii\hack@}

\def\markup#1{\let\@@marksf=\empty
     \ifhmode\edef\@@marksf{\spacefactor=\the\spacefactor\relax}\/\fi
     ${}^{\hbox{\subscriptfonts#1}}$\@@marksf}


\newtoks\shortyear
\newtoks\militaryhour
\newtoks\standardhour
\newtoks\minute
\newtoks\amorpm

\def\settime{\count@=\time\divide\count@ by60
     \militaryhour=\expandafter{\number\count@}%
     {\multiply\count@ by-60 \advance\count@ by\time
          \xdef\hack@{\ifnum\count@<10 0\fi\number\count@}}%
     \minute=\expandafter{\hack@}%
     \ifnum\count@<12
          \amorpm={am}
     \else\amorpm={pm}
          \ifnum\count@>12 \advance\count@ by-12 \fi
     \fi
     \standardhour=\expandafter{\number\count@}%
     \def\hack@19##1##2{\shortyear={##1##2}}%
          \expandafter\hack@\the\year}

\def\monthword#1{%
     \ifcase#1
          $\bullet$\err@badcountervalue{monthword}%
          \or January\or February\or March\or April\or May\or June%
          \or July\or August\or September\or October\or November\or December%
     \else$\bullet$\err@badcountervalue{monthword}%
     \fi}

\def\monthabbr#1{%
     \ifcase#1
          $\bullet$\err@badcountervalue{monthabbr}%
          \or Jan\or Feb\or Mar\or Apr\or May\or Jun%
          \or Jul\or Aug\or Sep\or Oct\or Nov\or Dec%
     \else$\bullet$\err@badcountervalue{monthabbr}%
     \fi}

\def\militarytime{\the\militaryhour:\the\minute}
\def\standardtime{\the\standardhour:\the\minute}


\def\@setnumstyle#1#2{\expandafter\global\expandafter\expandafter
     \expandafter\let\expandafter\expandafter
     \csname @\expandafter\blank\string#1style\endcsname
     \csname#2\endcsname}
\def\numstyle#1{\usename{@\expandafter\blank\string#1style}#1}
\def\ifblank#1\then{\useafter\ifx{@\expandafter\blank\string#1}\blank}

\def\blank#1{}

\def\Roman#1{\expandafter\uppercase\expandafter{\romannumeral#1}}
\def\alphabetic#1{%
     \ifcase#1
          $\bullet$\err@badcountervalue{alphabetic}%
          \or a\or b\or c\or d\or e\or f\or g\or h\or i\or j\or k\or l\or m%
          \or n\or o\or p\or q\or r\or s\or t\or u\or v\or w\or x\or y\or z%
     \else$\bullet$\err@badcountervalue{alphabetic}%
     \fi}
\def\Alphabetic#1{\expandafter\uppercase\expandafter{\alphabetic{#1}}}
\def\symbols#1{%
     \ifcase#1
          $\bullet$\err@badcountervalue{symbols}%
          \or*\or\dag\or\ddag\or\S\or$\|$%
          \or**\or\dag\dag\or\ddag\ddag\or\S\S\or$\|\|$%
     \else$\bullet$\err@badcountervalue{symbols}%
     \fi}


\catcode`\^^?=13 \def^^?{\relax}

\def\trimleading#1\to#2{\edef#2{#1}%
     \expandafter\@trimleading\expandafter#2#2^^?^^?}
\def\@trimleading#1#2#3^^?{\ifx#2^^?\def#1{}\else\def#1{#2#3}\fi}

\def\trimtrailing#1\to#2{\edef#2{#1}%
     \expandafter\@trimtrailing\expandafter#2#2^^? ^^?\relax}
\def\@trimtrailing#1#2 ^^?#3{\ifx#3\relax\toks@={}%
     \else\def#1{#2}\toks@={\trimtrailing#1\to#1}\fi
     \the\toks@}

\def\trim#1\to#2{\trimleading#1\to#2\trimtrailing#2\to#2}

\catcode`\^^?=15


\long\def\additemL#1\to#2{\toks@={\^^\{#1}}\toks@ii=\expandafter{#2}%
     \xdef#2{\the\toks@\the\toks@ii}}

\long\def\additemR#1\to#2{\toks@={\^^\{#1}}\toks@ii=\expandafter{#2}%
     \xdef#2{\the\toks@ii\the\toks@}}

\def\getitemL#1\to#2{\expandafter\@getitemL#1\hack@#1#2}
\def\@getitemL\^^\#1#2\hack@#3#4{\def#4{#1}\def#3{#2}}

\message{font macros,}


\newdimen\rp@
\newcount\@@sizeindex \@@sizeindex=0
\newcount\@@factori
\newcount\@@factorii
\newcount\@@factoriii
\newcount\@@factoriv

\countdef\maxfam=18
\newfam\itfam
\newfam\bffam
\newfam\bfsfam
\newfam\bmitfam

\def\@mathfontinit{\count@=4
     \loop\textfont\count@=\nullfont
          \scriptfont\count@=\nullfont
          \scriptscriptfont\count@=\nullfont
          \ifnum\count@<\maxfam\advance\count@ by\@ne
     \repeat}

\def\@fontstyleinit{%
     \def\it{\err@fontnotavailable\it}%
     \def\bf{\err@fontnotavailable\bf}%
     \def\bfs{\err@bfstobf}%
     \def\bmit{\err@fontnotavailable\bmit}%
     \def\sc{\err@fontnotavailable\sc}%
     \def\sl{\err@sltoit}%
     \def\ss{\err@fontnotavailable\ss}%
     \def\tt{\err@fontnotavailable\tt}}

\def\@parameterinit#1{\rm\rp@=.1em \@getscaling{#1}%
     \let\^^\=\@doscaling\scalingskipslist
     \setbox\strutbox=\hbox{\vrule
          height.708\baselineskip depth.292\baselineskip width\z@}}

\def\@getfactor#1#2#3#4{\@@factori=#1 \@@factorii=#2
     \@@factoriii=#3 \@@factoriv=#4}

\def\@getscaling#1{\count@=#1 \advance\count@ by-\@@sizeindex\@@sizeindex=#1
     \ifnum\count@<0
          \let\@mulordiv=\divide
          \let\@divormul=\multiply
          \multiply\count@ by\m@ne
     \else\let\@mulordiv=\multiply
          \let\@divormul=\divide
     \fi
     \edef\@@scratcha{\ifcase\count@                {1}{1}{1}{1}\or
          {1}{7}{23}{3}\or     {2}{5}{3}{1}\or      {9}{89}{13}{1}\or
          {6}{25}{6}{1}\or     {8}{71}{14}{1}\or    {6}{25}{36}{5}\or
          {1}{7}{53}{4}\or     {12}{125}{108}{5}\or {3}{14}{53}{5}\or
          {6}{41}{17}{1}\or    {13}{31}{13}{2}\or   {9}{107}{71}{2}\or
          {11}{139}{124}{3}\or {1}{6}{43}{2}\or     {10}{107}{42}{1}\or
          {1}{5}{43}{2}\or     {5}{69}{65}{1}\or    {11}{97}{91}{2}\fi}%
     \expandafter\@getfactor\@@scratcha}

\def\@doscaling#1{\@mulordiv#1by\@@factori\@divormul#1by\@@factorii
     \@mulordiv#1by\@@factoriii\@divormul#1by\@@factoriv}


\newskip\headskip
\newskip\footskip

\def\typesize=#1pt{\count@=#1 \advance\count@ by-10
     \ifcase\count@
          \@setsizex\or\err@badtypesize\or
          \@setsizexii\or\err@badtypesize\or
          \@setsizexiv
     \else\err@badtypesize
     \fi}

\def\@setsizex{\getixpt
     \def\subsubscriptfonts{\vpt}%
          \def\subsubscriptsize{\vpt\@parameterinit{-8}}%
     \def\subscriptfonts{\viipt}\def\subscriptsize{\viipt\@parameterinit{-4}}%
     \def\footnotefonts{\viiipt}\def\footnotesize{\viiipt\@parameterinit{-2}}%
     \def\smallfonts{\ixpt}\def\smallsize{\ixpt\@parameterinit{-1}}%
     \def\normalfonts{\xpt}\def\normalsize{\xpt\@parameterinit{0}}%
     \def\bigfonts{\xiipt}\def\bigsize{\xiipt\@parameterinit{2}}%
     \def\Bigfonts{\xivpt}\def\Bigsize{\xivpt\@parameterinit{4}}%
     \def\biggfonts{\xviipt}\def\biggsize{\xviipt\@parameterinit{6}}%
     \def\Biggfonts{\xxipt}\def\Biggsize{\xxipt\@parameterinit{8}}%
     \def\tinyfonts{\vpt}\def\tinysize{\vpt\@parameterinit{-8}}%
     \def\HUGEFONTS{\xxvpt}\def\HUGESIZE{\xxvpt\@parameterinit{10}}%
     \normalsize\fixedskipslist}

\def\@setsizexii{\getxipt
     \def\subsubscriptfonts{\vipt}%
          \def\subsubscriptsize{\vipt\@parameterinit{-6}}%
     \def\subscriptfonts{\viiipt}%
          \def\subscriptsize{\viiipt\@parameterinit{-2}}%
     \def\footnotefonts{\xpt}\def\footnotesize{\xpt\@parameterinit{0}}%
     \def\smallfonts{\xipt}\def\smallsize{\xipt\@parameterinit{1}}%
     \def\normalfonts{\xiipt}\def\normalsize{\xiipt\@parameterinit{2}}%
     \def\bigfonts{\xivpt}\def\bigsize{\xivpt\@parameterinit{4}}%
     \def\Bigfonts{\xviipt}\def\Bigsize{\xviipt\@parameterinit{6}}%
     \def\biggfonts{\xxipt}\def\biggsize{\xxipt\@parameterinit{8}}%
     \def\Biggfonts{\xxvpt}\def\Biggsize{\xxvpt\@parameterinit{10}}%
     \def\tinyfonts{\vpt}\def\tinysize{\vpt\@parameterinit{-8}}%
     \def\HUGEFONTS{\xxvpt}\def\HUGESIZE{\xxvpt\@parameterinit{10}}%
     \normalsize\fixedskipslist}

\def\@setsizexiv{\getxiiipt
     \def\subsubscriptfonts{\viipt}%
          \def\subsubscriptsize{\viipt\@parameterinit{-4}}%
     \def\subscriptfonts{\xpt}\def\subscriptsize{\xpt\@parameterinit{0}}%
     \def\footnotefonts{\xiipt}\def\footnotesize{\xiipt\@parameterinit{2}}%
     \def\smallfonts{\xiiipt}\def\smallsize{\xiiipt\@parameterinit{3}}%
     \def\normalfonts{\xivpt}\def\normalsize{\xivpt\@parameterinit{4}}%
     \def\bigfonts{\xviipt}\def\bigsize{\xviipt\@parameterinit{6}}%
     \def\Bigfonts{\xxipt}\def\Bigsize{\xxipt\@parameterinit{8}}%
     \def\biggfonts{\xxvpt}\def\biggsize{\xxvpt\@parameterinit{10}}%
     \def\Biggfonts{\err@sizetoolarge\Biggfonts\HUGEFONTS}%
          \def\Biggsize{\err@sizetoolarge\Biggsize\HUGESIZE}%
     \def\tinyfonts{\vpt}\def\tinysize{\vpt\@parameterinit{-8}}%
     \def\HUGEFONTS{\xxvpt}\def\HUGESIZE{\xxvpt\@parameterinit{10}}%
     \normalsize\fixedskipslist}

\def\subsubscriptfonts{\vpt} \def\subsubscriptsize{\vpt\@parameterinit{-8}}
\def\subscriptfonts{\viipt}  \def\subscriptsize{\viipt\@parameterinit{-4}}
\def\footnotefonts{\viiipt}  \def\footnotesize{\viiipt\@parameterinit{-2}}
\def\smallfonts{\err@sizenotavailable\smallfonts}
                             \def\smallsize{\ixpt\@parameterinit{-1}}
\def\normalfonts{\xpt}       \def\normalsize{\xpt\@parameterinit{0}}
\def\bigfonts{\xiipt}        \def\bigsize{\xiipt\@parameterinit{2}}
\def\Bigfonts{\xivpt}        \def\Bigsize{\xivpt\@parameterinit{4}}
\def\biggfonts{\xviipt}      \def\biggsize{\xviipt\@parameterinit{6}}
\def\Biggfonts{\xxipt}       \def\Biggsize{\xxipt\@parameterinit{8}}
\def\tinyfonts{\vpt}         \def\tinysize{\vpt\@parameterinit{-8}}
\def\HUGEFONTS{\xxvpt}       \def\HUGESIZE{\xxvpt\@parameterinit{10}}

\message{document layout,}


\newtoks\everyoutput \everyoutput={}
\newdimen\depthofpage
\newcount\pagenum \pagenum=0

\newdimen\oddtopmargin  \newdimen\eventopmargin
\newdimen\oddleftmargin \newdimen\evenleftmargin
\newtoks\oddhead        \newtoks\evenhead
\newtoks\oddfoot        \newtoks\evenfoot

\def\topmargin{\afterassignment\@seteventop\oddtopmargin}
\def\leftmargin{\afterassignment\@setevenleft\oddleftmargin}
\def\head{\afterassignment\@setevenhead\oddhead}
\def\foot{\afterassignment\@setevenfoot\oddfoot}

\def\@seteventop{\eventopmargin=\oddtopmargin}
\def\@setevenleft{\evenleftmargin=\oddleftmargin}
\def\@setevenhead{\evenhead=\oddhead}
\def\@setevenfoot{\evenfoot=\oddfoot}

\def\pagenumstyle#1{\@setnumstyle\pagenum{#1}}

\newif\ifdraft
\def\draft{\drafttrue\leftmargin=.5in \overfullrule=5pt }

\def\outputstyle#1{\global\expandafter\let\expandafter
          \@outputstyle\csname#1output\endcsname
     \usename{#1setup}}

\output={\@outputstyle}

\def\normaloutput{\the\everyoutput
     \global\advance\pagenum by\@ne
     \ifodd\pagenum
          \voffset=\oddtopmargin \hoffset=\oddleftmargin
     \else\voffset=\eventopmargin \hoffset=\evenleftmargin
     \fi
     \advance\voffset by-1in  \advance\hoffset by-1in
     \count0=\pagenum
     \expandafter\shipout\pagebox
     \ifnum\outputpenalty>-\@MM\else\dosupereject\fi}

\newdimen\fullhsize
\newbox\leftpage
\newcount\leftpagenum
\newcount\outputpagenum \outputpagenum=0
\let\leftorright=L

\def\twoupoutput{\the\everyoutput
     \global\advance\pagenum by\@ne
     \if L\leftorright
          \global\setbox\leftpage=\leftline{\pagebox}%
          \global\leftpagenum=\pagenum
          \global\let\leftorright=R%
     \else\global\advance\outputpagenum by\@ne
          \ifodd\outputpagenum
               \voffset=\oddtopmargin \hoffset=\oddleftmargin
          \else\voffset=\eventopmargin \hoffset=\evenleftmargin
          \fi
          \advance\voffset by-1in  \advance\hoffset by-1in
          \count0=\leftpagenum \count1=\pagenum
          \shipout\vbox{\hbox to\fullhsize
               {\box\leftpage\hfil\leftline{\pagebox}}}%
          \global\let\leftorright=L%
     \fi
     \ifnum\outputpenalty>-\@MM
     \else\dosupereject
          \if R\leftorright
               \globaldefs=\@ne\head={\hfil}\foot={\hfil}\globaldefs=\z@
               \null\newpage
          \fi
     \fi}

\def\pagebox{\vbox{\makeheadline\pagebody\makefootline}}

\def\makeheadline{%
     \vbox to\z@{\baselinestretch=\@m
          \vskip\topskip\vskip-.708\baselineskip\vskip-\headskip
          \line{\vbox to\ht\strutbox{}%
               \ifodd\pagenum\the\oddhead\else\the\evenhead\fi}%
          \vss}%
     \nointerlineskip}

\def\pagebody{\vbox to\vsize{%
     \boxmaxdepth\maxdepth
     \ifvoid\topins\else\unvbox\topins\fi
     \depthofpage=\dp255
     \unvbox255
     \ifraggedbottom\kern-\depthofpage\vfil\fi
     \ifvoid\footins
     \else\vskip\skip\footins
          \footnoterule
          \unvbox\footins
          \vskip-\footnoteskip
     \fi}}

\def\makefootline{\baselineskip=\footskip
     \line{\ifodd\pagenum\the\oddfoot\else\the\evenfoot\fi}}


\newskip\abovechapterskip
\newskip\belowchapterskip
\newskip\abovesectionskip
\newskip\belowsectionskip
\newskip\abovesubsectionskip
\newskip\belowsubsectionskip

\def\chapterstyle#1{\global\expandafter\let\expandafter\@chapterstyle
     \csname#1text\endcsname}
\def\sectionstyle#1{\global\expandafter\let\expandafter\@sectionstyle
     \csname#1text\endcsname}
\def\subsectionstyle#1{\global\expandafter\let\expandafter\@subsectionstyle
     \csname#1text\endcsname}

\def\chapter#1{%
     \ifdim\lastskip=17sp \else\chapterbreak\vskip\abovechapterskip\fi
     \@chapterstyle{\ifblank\chapternumstyle\then
          \else\newchapternum=\next\chapternumformat\ \fi#1}%
     \nobreak\vskip\belowchapterskip\vskip17sp }

\def\section#1{%
     \ifdim\lastskip=17sp \else\sectionbreak\vskip\abovesectionskip\fi
     \@sectionstyle{\ifblank\sectionnumstyle\then
          \else\newsectionnum=\next\sectionnumformat\ \fi#1}%
     \nobreak\vskip\belowsectionskip\vskip17sp }

\def\subsection#1{%
     \ifdim\lastskip=17sp \else\subsectionbreak\vskip\abovesubsectionskip\fi
     \@subsectionstyle{\ifblank\subsectionnumstyle\then
          \else\newsubsectionnum=\next\subsectionnumformat\ \fi#1}%
     \nobreak\vskip\belowsubsectionskip\vskip17sp }


\let\TeXunderline=\underline
\let\TeXoverline=\overline
\def\underline#1{\relax\ifmmode\TeXunderline{#1}\else
     $\TeXunderline{\hbox{#1}}$\fi}
\def\overline#1{\relax\ifmmode\TeXoverline{#1}\else
     $\TeXoverline{\hbox{#1}}$\fi}

\def\baselinestretch{\afterassignment\@baselinestretch\count@}
\def\@baselinestretch{\baselineskip=\normalbaselineskip
     \divide\baselineskip by\@m\baselineskip=\count@\baselineskip
     \setbox\strutbox=\hbox{\vrule
          height.708\baselineskip depth.292\baselineskip width\z@}%
     \bigskipamount=\the\baselineskip
          plus.25\baselineskip minus.25\baselineskip
     \medskipamount=.5\baselineskip
          plus.125\baselineskip minus.125\baselineskip
     \smallskipamount=.25\baselineskip
          plus.0625\baselineskip minus.0625\baselineskip}

\def\\{\ifhmode\ifnum\lastpenalty=-\@M\else\hfil\penalty-\@M\fi\fi
     \ignorespaces}
\def\newpage{\vfil\break}

\def\lefttext#1{\par{\@text\leftskip=\z@\rightskip=\centering
     \noindent#1\par}}
\def\righttext#1{\par{\@text\leftskip=\centering\rightskip=\z@
     \noindent#1\par}}
\def\centertext#1{\par{\@text\leftskip=\centering\rightskip=\centering
     \noindent#1\par}}
\def\@text{\parindent=\z@ \parfillskip=\z@ \everypar={}%
     \spaceskip=.3333em \xspaceskip=.5em
     \def\\{\ifhmode\ifnum\lastpenalty=-\@M\else\penalty-\@M\fi\fi
          \ignorespaces}}

\def\beginleft{\par\@text\leftskip=\z@ \rightskip=\centering}
     
\def\beginright{\par\@text\leftskip=\centering\rightskip=\z@ }
     
\def\begincenter{\par\@text\leftskip=\centering\rightskip=\centering}

\def\beginnarrow{\defaultoption[\parindent]\@beginnarrow}
\def\@beginnarrow[#1]{\par\advance\leftskip by#1\advance\rightskip by#1}

\begingroup
\catcode`\[=1 \catcode`\{=11 \gdef\beginignore[\endgroup\bgroup
     \catcode`\e=0 \catcode`\\=12 \catcode`\{=11 \catcode`\f=12 \let\or=\relax
     \let\nd{ignor=\fi \let\}=\egroup
     \iffalse}
\endgroup

\long\def\marginnote#1{\leavevmode
     \edef\@marginsf{\spacefactor=\the\spacefactor\relax}%
     \ifdraft\strut\vadjust{%
          \hbox to\z@{\hskip\hsize\hskip.1in
               \vbox to\z@{\vskip-\dp\strutbox
                    \marginnoteformat
                    \vskip-\ht\strutbox
                    \noindent\strut#1\par
                    \vss}%
               \hss}}%
     \fi
     \@marginsf}


\newtoks\everybye \everybye={\par\vfil}
\outer\def\bye{\the\everybye
     \footnotecheck
     \prelabelcheck
     \streamcheck
     \supereject
     \TeXend}

\message{footnotes,}

\newcount\footnotenum \footnotenum=0
\newskip\footnoteskip
\let\@footnotelist=\empty

\def\footnotenumstyle#1{\@setnumstyle\footnotenum{#1}%
     \useafter\ifx{@footnotenumstyle}\symbols
          \global\let\@footup=\empty
     \else\global\let\@footup=\markup
     \fi}

\def\footnote{\footnotecheck\defaultoption[]\@footnote}
\def\@footnote[#1]{\@footnotemark[#1]\@footnotetext}

\def\footnotemark{\defaultoption[]\@footnotemark}
\def\@footnotemark[#1]{\let\@footsf=\empty
     \ifhmode\edef\@footsf{\spacefactor=\the\spacefactor\relax}\/\fi
     \ifnoarg#1\then
          \global\advance\footnotenum by\@ne
          \@footup{\footnotenumformat}%
          \edef\@@foota{\footnotenum=\the\footnotenum\relax}%
          \expandafter\additemR\expandafter\@footup\expandafter
               {\@@foota\footnotenumformat}\to\@footnotelist
          \global\let\@footnotelist=\@footnotelist
     \else\markup{#1}%
          \additemR\markup{#1}\to\@footnotelist
          \global\let\@footnotelist=\@footnotelist
     \fi
     \@footsf}

\def\footnotetext{%
     \ifx\@footnotelist\empty\err@extrafootnotetext\else\@footnotetext\fi}
\def\@footnotetext{%
     \getitemL\@footnotelist\to\@@foota
     \global\let\@footnotelist=\@footnotelist
     \insert\footins\bgroup
     \footnoteformat
     \splittopskip=\ht\strutbox\splitmaxdepth=\dp\strutbox
     \interlinepenalty=\interfootnotelinepenalty\floatingpenalty=\@MM
     \noindent\llap{\@@foota}\strut
     \bgroup\aftergroup\@footnoteend
     \let\@@scratcha=}
\def\@footnoteend{\strut\par\vskip\footnoteskip\egroup}

\def\footnoterule{\normalfonts
     \kern-.3em \hrule width2in height.04em \kern .26em }

\def\footnotecheck{%
     \ifx\@footnotelist\empty
     \else\err@extrafootnotemark
          \global\let\@footnotelist=\empty
     \fi}

\message{labels,}

\let\@@labeldef=\xdef
\newif\if@labelfile
\newwrite\@labelfile
\let\@prelabellist=\empty

\def\label#1#2{\trim#1\to\@@labarg\edef\@@labtext{#2}%
     \edef\@@labname{lab@\@@labarg}%
     \useafter\ifundefined\@@labname\then\else\@yeslab\fi
     \useafter\@@labeldef\@@labname{#2}%
     \ifstreaming
          \expandafter\toks@\expandafter\expandafter\expandafter
               {\csname\@@labname\endcsname}%
          \immediate\write\streamout{\noexpand\label{\@@labarg}{\the\toks@}}%
     \fi}
\def\@yeslab{%
     \useafter\ifundefined{if\@@labname}\then
          \err@labelredef\@@labarg
     \else\useif{if\@@labname}\then
               \err@labelredef\@@labarg
          \else\global\usename{\@@labname true}%
               \useafter\ifundefined{pre\@@labname}\then
               \else\useafter\ifx{pre\@@labname}\@@labtext
                    \else\err@badlabelmatch\@@labarg
                    \fi
               \fi
               \if@labelfile
               \else\global\@labelfiletrue
                    \immediate\write\sixt@@n{--> Creating file \jobname.lab}%
                    \immediate\openout\@labelfile=\jobname.lab
               \fi
               \immediate\write\@labelfile
                    {\noexpand\prelabel{\@@labarg}{\@@labtext}}%
          \fi
     \fi}

\def\putlab#1{\trim#1\to\@@labarg\edef\@@labname{lab@\@@labarg}%
     \useafter\ifundefined\@@labname\then\@nolab\else\usename\@@labname\fi}
\def\@nolab{%
     \useafter\ifundefined{pre\@@labname}\then
          \undefinedlabelformat
          \err@needlabel\@@labarg
          \useafter\xdef\@@labname{\undefinedlabelformat}%
     \else\usename{pre\@@labname}%
          \useafter\xdef\@@labname{\usename{pre\@@labname}}%
     \fi
     \useafter\newif{if\@@labname}%
     \expandafter\additemR\@@labarg\to\@prelabellist}

\def\prelabel#1{\useafter\gdef{prelab@#1}}

\def\ifundefinedlabel#1\then{%
     \expandafter\ifx\csname lab@#1\endcsname\relax}
\def\useiflab#1\then{\csname iflab@#1\endcsname}

\def\prelabelcheck{{%
     \def\^^\##1{\useiflab{##1}\then\else\err@undefinedlabel{##1}\fi}%
     \@prelabellist}}

\message{equation numbering,}

\newcount\chapternum
\newcount\sectionnum
\newcount\subsectionnum
\newcount\equationnum
\newcount\subequationnum
\newcount\figurenum
\newcount\subfigurenum
\newcount\tablenum
\newcount\subtablenum

\newif\if@subeqncount
\newif\if@subfigcount
\newif\if@subtblcount

\def\newchapternum{\newsectionnum=\z@\@resetnum\chapternum}
\def\newsectionnum{\newsubsectionnum=\z@\@resetnum\sectionnum}
\def\newsubsectionnum{\newequationnum=\z@\newfigurenum=\z@\newtablenum=\z@
     \@resetnum\subsectionnum}
\def\newequationnum{\newsubequationnum=\z@\@resetnum\equationnum}
\def\newsubequationnum{\@resetnum\subequationnum}
\def\newfigurenum{\newsubfigurenum=\z@\@resetnum\figurenum}
\def\newsubfigurenum{\@resetnum\subfigurenum}
\def\newtablenum{\newsubtablenum=\z@\@resetnum\tablenum}
\def\newsubtablenum{\@resetnum\subtablenum}

\def\@resetnum#1{\global\advance#1by1 \edef\next{\the#1\relax}\global#1}

\newchapternum=0

\def\chapternumstyle#1{\@setnumstyle\chapternum{#1}}
\def\sectionnumstyle#1{\@setnumstyle\sectionnum{#1}}
\def\subsectionnumstyle#1{\@setnumstyle\subsectionnum{#1}}
\def\equationnumstyle#1{\@setnumstyle\equationnum{#1}}
\def\subequationnumstyle#1{\@setnumstyle\subequationnum{#1}%
     \ifblank\subequationnumstyle\then\global\@subeqncountfalse\fi
     \ignorespaces}
\def\figurenumstyle#1{\@setnumstyle\figurenum{#1}}
\def\subfigurenumstyle#1{\@setnumstyle\subfigurenum{#1}%
     \ifblank\subfigurenumstyle\then\global\@subfigcountfalse\fi
     \ignorespaces}
\def\tablenumstyle#1{\@setnumstyle\tablenum{#1}}
\def\subtablenumstyle#1{\@setnumstyle\subtablenum{#1}%
     \ifblank\subtablenumstyle\then\global\@subtblcountfalse\fi
     \ignorespaces}

\def\eqnlabel#1{%
     \if@subeqncount
          \newsubequationnum=\next
     \else\newequationnum=\next
          \ifblank\subequationnumstyle\then
          \else\global\@subeqncounttrue
               \newsubequationnum=\@ne
          \fi
     \fi
     \label{#1}{\puteqnformat}(\puteqn{#1})%
     \ifdraft\rlap{\hskip.1in{\tt#1}}\fi}

\let\puteqn=\putlab

\def\equation#1#2{\useafter\gdef{eqn@#1}{#2\eqno\eqnlabel{#1}}}
\def\Equation#1{\useafter\gdef{eqn@#1}}

\def\putequation#1{\useafter\ifundefined{eqn@#1}\then
     \err@undefinedeqn{#1}\else\usename{eqn@#1}\fi}

\def\eqnseriesstyle#1{\gdef\@eqnseriesstyle{#1}}
\def\begineqnseries{\subequationnumstyle{\@eqnseriesstyle}%
     \defaultoption[]\@begineqnseries}
\def\@begineqnseries[#1]{\edef\@@eqnname{#1}}
\def\endeqnseries{\subequationnumstyle{blank}%
     \expandafter\ifnoarg\@@eqnname\then
     \else\label\@@eqnname{\puteqnformat}%
     \fi
     \aftergroup\ignorespaces}

\def\figlabel#1{%
     \if@subfigcount
          \newsubfigurenum=\next
     \else\newfigurenum=\next
          \ifblank\subfigurenumstyle\then
          \else\global\@subfigcounttrue
               \newsubfigurenum=\@ne
          \fi
     \fi
     \label{#1}{\putfigformat}\putfig{#1}%
     {\def\marginnoteformat{\tt}\marginnote{#1}}}

\let\putfig=\putlab

\def\figseriesstyle#1{\gdef\@figseriesstyle{#1}}
\def\beginfigseries{\subfigurenumstyle{\@figseriesstyle}%
     \defaultoption[]\@beginfigseries}
\def\@beginfigseries[#1]{\edef\@@figname{#1}}
\def\endfigseries{\subfigurenumstyle{blank}%
     \expandafter\ifnoarg\@@figname\then
     \else\label\@@figname{\putfigformat}%
     \fi
     \aftergroup\ignorespaces}

\def\tbllabel#1{%
     \if@subtblcount
          \newsubtablenum=\next
     \else\newtablenum=\next
          \ifblank\subtablenumstyle\then
          \else\global\@subtblcounttrue
               \newsubtablenum=\@ne
          \fi
     \fi
     \label{#1}{\puttblformat}\puttbl{#1}%
     {\def\marginnoteformat{\tt}\marginnote{#1}}}

\let\puttbl=\putlab

\def\tblseriesstyle#1{\gdef\@tblseriesstyle{#1}}
\def\begintblseries{\subtablenumstyle{\@tblseriesstyle}%
     \defaultoption[]\@begintblseries}
\def\@begintblseries[#1]{\edef\@@tblname{#1}}
\def\endtblseries{\subtablenumstyle{blank}%
     \expandafter\ifnoarg\@@tblname\then
     \else\label\@@tblname{\puttblformat}%
     \fi
     \aftergroup\ignorespaces}

\message{reference numbering,}

\newcount\referencenum \referencenum=0
\newcount\@@prerefcount \@@prerefcount=0
\newcount\@@thisref
\newcount\@@lastref
\newcount\@@loopref
\newcount\@@refseq
\newdimen\refnumindent
\let\@undefreflist=\empty

\def\referencenumstyle#1{\@setnumstyle\referencenum{#1}}

\def\referencestyle#1{\usename{@ref#1}}

\def\@refsequential{%
     \gdef\@refpredef##1{\global\advance\referencenum by\@ne
          \let\^^\=0\label{##1}{\^^\{\the\referencenum}}%
          \useafter\gdef{ref@\the\referencenum}{{##1}{\undefinedlabelformat}}}%
     \gdef\@reference##1##2{%
          \ifundefinedlabel##1\then
          \else\def\^^\####1{\global\@@thisref=####1\relax}\putlab{##1}%
               \useafter\gdef{ref@\the\@@thisref}{{##1}{##2}}%
          \fi}%
     \gdef\endputreferences{%
          \loop\ifnum\@@loopref<\referencenum
                    \advance\@@loopref by\@ne
                    \expandafter\expandafter\expandafter\@printreference
                         \csname ref@\the\@@loopref\endcsname
          \repeat
          \par}}

\def\@refpreordered{%
     \gdef\@refpredef##1{\global\advance\referencenum by\@ne
          \additemR##1\to\@undefreflist}%
     \gdef\@reference##1##2{%
          \ifundefinedlabel##1\then
          \else\global\advance\@@loopref by\@ne
               {\let\^^\=0\label{##1}{\^^\{\the\@@loopref}}}%
               \@printreference{##1}{##2}%
          \fi}
     \gdef\endputreferences{%
          \def\^^\####1{\useiflab{####1}\then
               \else\reference{####1}{\undefinedlabelformat}\fi}%
          \@undefreflist
          \par}}

\def\beginprereferences{\par
     \def\reference##1##2{\global\advance\referencenum by1\@ne
          \let\^^\=0\label{##1}{\^^\{\the\referencenum}}%
          \useafter\gdef{ref@\the\referencenum}{{##1}{##2}}}}
\def\endprereferences{\global\@@prerefcount=\the\referencenum\par}

\def\beginputreferences{\par
     \refnumindent=\z@\@@loopref=\z@
     \loop\ifnum\@@loopref<\referencenum
               \advance\@@loopref by\@ne
               \setbox\z@=\hbox{\referencenum=\@@loopref
                    \referencenumformat\enskip}%
               \ifdim\wd\z@>\refnumindent\refnumindent=\wd\z@\fi
     \repeat
     \putreferenceformat
     \@@loopref=\z@
     \loop\ifnum\@@loopref<\@@prerefcount
               \advance\@@loopref by\@ne
               \expandafter\expandafter\expandafter\@printreference
                    \csname ref@\the\@@loopref\endcsname
     \repeat
     \let\reference=\@reference}

\def\@printreference#1#2{\ifx#2\undefinedlabelformat\err@undefinedref{#1}\fi
     \noindent\ifdraft\rlap{\hskip\hsize\hskip.1in \tt#1}\fi
     \llap{\referencenum=\@@loopref\referencenumformat\enskip}#2\par}

\def\reference#1#2{{\par\refnumindent=\z@\putreferenceformat\noindent#2\par}}

\def\putref#1{\trim#1\to\@@refarg
     \expandafter\ifnoarg\@@refarg\then
          \toks@={\relax}%
     \else\@@lastref=-\@m\def\@@refsep{}\def\@more{\@nextref}%
          \toks@={\@nextref#1,,}%
     \fi\the\toks@}
\def\@nextref#1,{\trim#1\to\@@refarg
     \expandafter\ifnoarg\@@refarg\then
          \let\@more=\relax
     \else\ifundefinedlabel\@@refarg\then
               \expandafter\@refpredef\expandafter{\@@refarg}%
          \fi
          \def\^^\##1{\global\@@thisref=##1\relax}%
          \global\@@thisref=\m@ne
          \setbox\z@=\hbox{\putlab\@@refarg}%
     \fi
     \advance\@@lastref by\@ne
     \ifnum\@@lastref=\@@thisref\advance\@@refseq by\@ne\else\@@refseq=\@ne\fi
     \ifnum\@@lastref<\z@
     \else\ifnum\@@refseq<\thr@@
               \@@refsep\def\@@refsep{,}%
               \ifnum\@@lastref>\z@
                    \advance\@@lastref by\m@ne
                    {\referencenum=\@@lastref\putrefformat}%
               \else\undefinedlabelformat
               \fi
          \else\def\@@refsep{--}%
          \fi
     \fi
     \@@lastref=\@@thisref
     \@more}

\message{streaming,}

\newif\ifstreaming

\def\streamto{\defaultoption[\jobname]\@streamto}
\def\@streamto[#1]{\global\streamingtrue
     \immediate\write\sixt@@n{--> Streaming to #1.str}%
     \newwrite\streamout\immediate\openout\streamout=#1.str }

\def\streamfrom{\defaultoption[\jobname]\@streamfrom}
\def\@streamfrom[#1]{\newread\streamin\openin\streamin=#1.str
     \ifeof\streamin
          \expandafter\err@nostream\expandafter{#1.str}%
     \else\immediate\write\sixt@@n{--> Streaming from #1.str}%
          \let\@@labeldef=\gdef
          \ifstreaming
               \edef\@elc{\endlinechar=\the\endlinechar}%
               \endlinechar=\m@ne
               \loop\read\streamin to\@@scratcha
                    \ifeof\streamin
                         \streamingfalse
                    \else\toks@=\expandafter{\@@scratcha}%
                         \immediate\write\streamout{\the\toks@}%
                    \fi
                    \ifstreaming
               \repeat
               \@elc
               \input #1.str
               \streamingtrue
          \else\input #1.str
          \fi
          \let\@@labeldef=\xdef
     \fi}

\def\streamcheck{\ifstreaming
     \immediate\write\streamout{\pagenum=\the\pagenum}%
     \immediate\write\streamout{\footnotenum=\the\footnotenum}%
     \immediate\write\streamout{\referencenum=\the\referencenum}%
     \immediate\write\streamout{\chapternum=\the\chapternum}%
     \immediate\write\streamout{\sectionnum=\the\sectionnum}%
     \immediate\write\streamout{\subsectionnum=\the\subsectionnum}%
     \immediate\write\streamout{\equationnum=\the\equationnum}%
     \immediate\write\streamout{\subequationnum=\the\subequationnum}%
     \immediate\write\streamout{\figurenum=\the\figurenum}%
     \immediate\write\streamout{\subfigurenum=\the\subfigurenum}%
     \immediate\write\streamout{\tablenum=\the\tablenum}%
     \immediate\write\streamout{\subtablenum=\the\subtablenum}%
     \immediate\closeout\streamout
     \fi}


\def\err@badtypesize{%
     \errhelp={The limited availability of certain fonts requires^^J%
          that the base type size be 10pt, 12pt, or 14pt.^^J}%
     \errmessage{--> Illegal base type size}}

\def\err@badsizechange{\immediate\write\sixt@@n
     {--> Size change not allowed in math mode, ignored}}

\def\err@sizetoolarge#1{\immediate\write\sixt@@n
     {--> \noexpand#1 too big, substituting HUGE}}

\def\err@sizenotavailable#1{\immediate\write\sixt@@n
     {--> Size not available, \noexpand#1 ignored}}

\def\err@fontnotavailable#1{\immediate\write\sixt@@n
     {--> Font not available, \noexpand#1 ignored}}

\def\err@sltoit{\immediate\write\sixt@@n
     {--> Style \noexpand\sl not available, substituting \noexpand\it}%
     \it}

\def\err@bfstobf{\immediate\write\sixt@@n
     {--> Style \noexpand\bfs not available, substituting \noexpand\bf}%
     \bf}

\def\err@badgroup#1#2{%
     \errhelp={The block you have just tried to close was not the one^^J%
          most recently opened.^^J}%
     \errmessage{--> \noexpand\end{#1} doesn't match \noexpand\begin{#2}}}

\def\err@badcountervalue#1{\immediate\write\sixt@@n
     {--> Counter (#1) out of bounds}}

\def\err@extrafootnotemark{\immediate\write\sixt@@n
     {--> \noexpand\footnotemark command
          has no corresponding \noexpand\footnotetext}}

\def\err@extrafootnotetext{%
     \errhelp{You have given a \noexpand\footnotetext command without first
          specifying^^Ja \noexpand\footnotemark.^^J}%
     \errmessage{--> \noexpand\footnotetext command has no corresponding
          \noexpand\footnotemark}}

\def\err@labelredef#1{\immediate\write\sixt@@n
     {--> Label "#1" redefined}}

\def\err@badlabelmatch#1{\immediate\write\sixt@@n
     {--> Definition of label "#1" doesn't match value in \jobname.lab}}

\def\err@needlabel#1{\immediate\write\sixt@@n
     {--> Label "#1" cited before its definition}}

\def\err@undefinedlabel#1{\immediate\write\sixt@@n
     {--> Label "#1" cited but never defined}}

\def\err@undefinedeqn#1{\immediate\write\sixt@@n
     {--> Equation "#1" not defined}}

\def\err@undefinedref#1{\immediate\write\sixt@@n
     {--> Reference "#1" not defined}}

\def\err@nostream#1{%
     \errhelp={You have tried to input a stream file that doesn't exist.^^J}%
     \errmessage{--> Stream file #1 not found}}

\message{jyTeX initialization}

\everyjob{\immediate\write16{--> jyTeX version \fmtversion}%
     \edef\@@jobname{\jobname}%
     \edef\jobname{\@@jobname}%
     \settime
     \openin0=\jobname.lab
     \ifeof0
     \else\closein0
          \immediate\write16{--> Getting labels from file \jobname.lab}%
          \input\jobname.lab
     \fi}


\def\fixedskipslist{%
     \^^\{\topskip}%
     \^^\{\splittopskip}%
     \^^\{\maxdepth}%
     \^^\{\skip\topins}%
     \^^\{\skip\footins}%
     \^^\{\headskip}%
     \^^\{\footskip}}

\def\scalingskipslist{%
     \^^\{\p@renwd}%
     \^^\{\delimitershortfall}%
     \^^\{\nulldelimiterspace}%
     \^^\{\scriptspace}%
     \^^\{\jot}%
     \^^\{\normalbaselineskip}%
     \^^\{\normallineskip}%
     \^^\{\normallineskiplimit}%
     \^^\{\baselineskip}%
     \^^\{\lineskip}%
     \^^\{\lineskiplimit}%
     \^^\{\bigskipamount}%
     \^^\{\medskipamount}%
     \^^\{\smallskipamount}%
     \^^\{\parskip}%
     \^^\{\parindent}%
     \^^\{\abovedisplayskip}%
     \^^\{\belowdisplayskip}%
     \^^\{\abovedisplayshortskip}%
     \^^\{\belowdisplayshortskip}%
     \^^\{\abovechapterskip}%
     \^^\{\belowchapterskip}%
     \^^\{\abovesectionskip}%
     \^^\{\belowsectionskip}%
     \^^\{\abovesubsectionskip}%
     \^^\{\belowsubsectionskip}}


\def\twoupsetup{
     \topmargin=.75in
     \leftmargin=.5in
     \vsize=6.9in
     \hsize=4.75in
     \fullhsize=10in
     \let\draft=\relax}

\outputstyle{normal}                             

\def\marginnoteformat{\subscriptsize             
     \hsize=1in \baselinestretch=1000 \everypar={}%
     \tolerance=5000 \hbadness=5000 \parskip=0pt \parindent=0pt
     \leftskip=0pt \rightskip=0pt \raggedright}

\head={\ifdraft\normalfonts\it\hfil DRAFT\hfil   
     \llap{\number\day\ \monthword\month\ \militarytime}\else\hfil\fi}
\foot={\hfil\normalfonts\numstyle\pagenum\hfil}  

\normalbaselineskip=12pt                         
\normallineskip=0pt                              
\normallineskiplimit=0pt                         
\normalbaselines                                 

\topskip=.85\baselineskip \splittopskip=\topskip \headskip=2\baselineskip
\footskip=\headskip

\pagenumstyle{arabic}                            

\parskip=0pt                                     
\parindent=20pt                                  

\baselinestretch=1000                            


\chapterstyle{left}                              
\chapternumstyle{blank}                          
\def\chapterbreak{\newpage}                      
\abovechapterskip=0pt                            
\belowchapterskip=1.5\baselineskip               
     plus.38\baselineskip minus.38\baselineskip
\def\chapternumformat{\numstyle\chapternum.}     

\sectionstyle{left}                              
\sectionnumstyle{blank}                          
\def\sectionbreak{\vskip0pt plus4\baselineskip\penalty-100
     \vskip0pt plus-4\baselineskip}              
\abovesectionskip=1.5\baselineskip               
     plus.38\baselineskip minus.38\baselineskip
\belowsectionskip=\the\baselineskip              
     plus.25\baselineskip minus.25\baselineskip
\def\sectionnumformat{
     \ifblank\chapternumstyle\then\else\numstyle\chapternum.\fi
     \numstyle\sectionnum.}

\subsectionstyle{left}                           
\subsectionnumstyle{blank}                       
\def\subsectionbreak{\vskip0pt plus4\baselineskip\penalty-100
     \vskip0pt plus-4\baselineskip}              
\abovesubsectionskip=\the\baselineskip           
     plus.25\baselineskip minus.25\baselineskip
\belowsubsectionskip=.75\baselineskip            
     plus.19\baselineskip minus.19\baselineskip
\def\subsectionnumformat{
     \ifblank\chapternumstyle\then\else\numstyle\chapternum.\fi
     \ifblank\sectionnumstyle\then\else\numstyle\sectionnum.\fi
     \numstyle\subsectionnum.}


\footnotenumstyle{symbols}                       
\footnoteskip=0pt                                
\def\footnotenumformat{\numstyle\footnotenum}    
\def\footnoteformat{\footnotesize                
     \everypar={}\parskip=0pt \parfillskip=0pt plus1fil
     \leftskip=1em \rightskip=0pt
     \spaceskip=0pt \xspaceskip=0pt
     \def\\{\ifhmode\ifnum\lastpenalty=-10000
          \else\hfil\penalty-10000 \fi\fi\ignorespaces}}


\def\undefinedlabelformat{$\bullet$}             


\equationnumstyle{arabic}                        
\subequationnumstyle{blank}                      
\figurenumstyle{arabic}                          
\subfigurenumstyle{blank}                        
\tablenumstyle{arabic}                           
\subtablenumstyle{blank}                         

\eqnseriesstyle{alphabetic}                      
\figseriesstyle{alphabetic}                      
\tblseriesstyle{alphabetic}                      

\def\puteqnformat{\hbox{
     \ifblank\chapternumstyle\then\else\numstyle\chapternum.\fi
     \ifblank\sectionnumstyle\then\else\numstyle\sectionnum.\fi
     \ifblank\subsectionnumstyle\then\else\numstyle\subsectionnum.\fi
     \numstyle\equationnum
     \numstyle\subequationnum}}
\def\putfigformat{\hbox{
     \ifblank\chapternumstyle\then\else\numstyle\chapternum.\fi
     \ifblank\sectionnumstyle\then\else\numstyle\sectionnum.\fi
     \ifblank\subsectionnumstyle\then\else\numstyle\subsectionnum.\fi
     \numstyle\figurenum
     \numstyle\subfigurenum}}
\def\puttblformat{\hbox{
     \ifblank\chapternumstyle\then\else\numstyle\chapternum.\fi
     \ifblank\sectionnumstyle\then\else\numstyle\sectionnum.\fi
     \ifblank\subsectionnumstyle\then\else\numstyle\subsectionnum.\fi
     \numstyle\tablenum
     \numstyle\subtablenum}}


\referencestyle{sequential}                      
\referencenumstyle{arabic}                       
\def\putrefformat{\numstyle\referencenum}        
\def\referencenumformat{\numstyle\referencenum.} 
\def\putreferenceformat{
     \everypar={\hangindent=1em \hangafter=1 }%
     \def\\{\hfil\break\null\hskip-1em \ignorespaces}%
     \leftskip=\refnumindent\parindent=0pt \interlinepenalty=1000 }


\normalsize


\def\fmtversion{2.6M (June 1992)}

\catcode`\@=12

\typesize=10pt \magnification=1200 \baselineskip17truept
\footnotenumstyle{arabic} \hsize=6truein\vsize=8.5truein
\input epsf
\sectionnumstyle{blank}
\chapternumstyle{blank}
\chapternum=1
\sectionnum=1
\pagenum=0

\def\begintitle{\pagenumstyle{blank}\parindent=0pt
\begin{narrow}[0.4in]}
\def\endtitle{\end{narrow}\newpage\pagenumstyle{arabic}}


\def\beginexercise{\vskip 20truept\parindent=0pt\begin{narrow}[10
truept]}
\def\endexercise{\vskip 10truept\end{narrow}}


\def\eql#1{\eqno\eqnlabel{#1}}
\def\ref{\reference}
\def\peq{\puteqn}
\def\pref{\putref}

\def\mgn{\marginnote}
\def\bex{\begin{exercise}}
\def\eex{\end{exercise}}


\font\open=msbm10 


\def\StretchRtArr#1{{\count255=0\loop\relbar\joinrel\advance\count255 by1
\ifnum\count255<#1\repeat\rightarrow}}
\def\StretchLtArr#1{\,{\leftarrow\!\!\count255=0\loop\relbar
\joinrel\advance\count255 by1\ifnum\count255<#1\repeat}}

\def\StretchLRtArr#1{\,{\leftarrow\!\!\count255=0\loop\relbar\joinrel\advance
\count255 by1\ifnum\count255<#1\repeat\rightarrow\,\,}}

\def\mbox#1{{\leavevmode\hbox{#1}}}

\def\hspace#1{{\phantom{\mbox#1}}}
\def\oZ{\mbox{\open\char90}}

\def\al{\alpha}
\def\bom{{\bmit\omega}}

\def\ga{\gamma}
\def\de{\delta}
\def\Ga{\Gamma}

\def\la{\lambda}
\def\La{\Lambda}
\def\om{\omega}

\def\th{\theta}

\def\ze{\zeta}

\def\De{\Delta}

\def\caC{{\cal C}}

\def\caO{{\cal O}}

\def\caS{{\cal S}}

\def\det{{\rm det\,}}

\def\sc{{\rm sc }}

\def\zf{$\zeta$--function}
\def\zfs{$\zeta$--functions}


\def\frac#1/#2{\leavevmode\kern.1em
\raise.5ex\hbox{\the\scriptfont0 #1}\kern-.1em/\kern-.15em
\lower.25ex\hbox{\the\scriptfont0 #2}}
\def\sfrac#1/#2{\leavevmode\kern.1em
\raise.5ex\hbox{\the\scriptscriptfont0 #1}\kern-.1em/\kern-.15em
\lower.25ex\hbox{\the\scriptscriptfont0 #2}}

\def\gtorder{\mathrel{\raise.3ex\hbox{$>$}\mkern-14mu
             \lower0.6ex\hbox{$\sim$}}}
\def\ltorder{\mathrel{\raise.3ex\hbox{$<$}\mkern-14mu
             \lower0.6ex\hbox{$\sim$}}}

\def\semidirprod{\rlap{\ss C}\raise1pt\hbox{$\mkern.75mu\times$}}
\def\for{\lower6pt\hbox{$\Big|$}}
\def\fish{\kern-.25em{\phantom{abcde}\over \phantom{abcde}}\kern-.25em}


\def\boxit#1{\vbox{\hrule\hbox{\vrule\kern3pt
        \vbox{\kern3pt#1\kern3pt}\kern3pt\vrule}\hrule}}
\def\dalemb#1#2{{\vbox{\hrule height .#2pt
        \hbox{\vrule width.#2pt height#1pt \kern#1pt \vrule
                width.#2pt} \hrule height.#2pt}}}

\def\ol{\overline}
\def\frac#1#2{{{#1}\over{#2}}}

\def\noin{\noindent}

\def\comb#1#2{{\left(#1\atop#2\right)}}

\def\nsl{\nabla\!\!\!\! / }

\def\cosech{{\rm cosech\,}}
\def\sech{{\rm sech\,}}

\def\eg{{\it e.g.}}
\def\ie{{\it i.e. }}
\def\cf{{\it cf }}

\def\av#1{\langle#1\rangle} 


\def\3j#1#2#3#4#5#6{\left\lgroup\matrix{#1&#2&#3\cr#4&#5&#6\cr}
\right\rgroup}

\def\m?{\mgn{?}}

\def\beq{\begin{eqnarray}}
\def\eeq{\end{eqnarray}}


\def\cmp#1#2#3{{\it Comm. Math. Phys.} {\bf {#1}} ({#2}) #3}
\def\cqg#1#2#3{{\it Class. Quant. Grav.} {\bf {#1}} ({#2}) #3}

\def\jpa#1#2#3{{\it J. Phys.} {\bf A{#1}} ({#2}) #3}

\def\np#1#2#3{{\it Nucl. Phys.} {\bf B{#1}} ({#2}) #3}

\def\pl#1#2#3{{\it Phys. Lett.} {\bf {#1}} ({#2}) #3}

\def\prD#1#2#3{{\it Phys. Rev.} {\bf D{#1}} ({#2}) #3}

\def\dmj#1#2#3{{\it Duke Math. J.} {\bf {#1}} ({#2}) #3}

\def\jpamt#1#2#3{{\it J. Phys.A:Math.Theor.} {\bf{#1}} ({#2}) #3}

\begin{title}
\vglue 0.5truein
\vskip15truept
\centertext {\Bigfonts \bf  Spherical Dirac GJMS } \vskip7truept
\vskip10truept\centertext{\Bigfonts \bf operator determinants }
 \vskip7truept
\vskip10truept\centertext{\Bigfonts \bf }
 \vskip 20truept
\centertext{J.S.Dowker\footnote{dowker@man.ac.uk;  dowkeruk@yahoo.co.uk}} \vskip
7truept \centertext{\it Theory Group,} \centertext{\it School of Physics and Astronomy,}
\centertext{\it The University of Manchester,} \centertext{\it Manchester, England} \vskip
7truept \centertext{}

\vskip 7truept \vskip40truept
\begin{narrow}
Motivated by AdS/CFT, the extension is made to spin--half of a scalar calculation of the
conformal anomalies and functional determinants of GJMS operators on spheres. The formal
aspects are heuristic but sufficient. A Barnes zeta--function representation again proves
effective.

The determinants are calculated for the two factorisations of the general $\Ga$--function
(intertwiner) form of the GJMS operator, and shown to be equal, even including any
multiplicative anomaly. A comment is made on the general eigenvalue problem and a few
numerical results are presented.

An alternative approach is detailed for odd dimensions and it is shown that the scalar
determinants are expressed in terms of the spinor ones, and {\it vice versa}. An explicit,
general form is given.
\end{narrow}
\vskip 5truept
\vskip 60truept
\vfil
\end{title}
\pagenum=0
\newpage

\section{\bf 1. Introduction}

The Dirac eigenproblem on spheres is a standard spectral topic. In particular cases,
cosmologically motivated, it dates back to the 1930s. It has occurred in quantum field
theory notions such as the conformal anomaly, effective action and the Casimir effect in
Kaluza--Klein geometries but, more lately, in connection with the AdS/CFT correspondence,
where spheres can be both boundary and dual bulk. It is the latter topic that provides my
motivation for the calculation that follows. Some of the results are known, in one form or
another, but I hope that the procedures here will be viewed as relatively rapid, and
sometimes more explicit than existing ones.

The idea is to extend the GJMS operator (on spheres) to spinors and to evaluate various
spectral invariants which have field theory significance. Although I do not discuss physics
here, motivation for the analysis can be found in aspects of the AdS/CFT correspondance in a
particular version of which the propagating operators take on a product form similar to the
GJMS operator, [\pref{Tseytlin}]. I also refer to the papers [\pref{AaandD,DandD,Diaz}]
for a more physics oriented perspective and I give some relevant remarks in section 10.

It is anticipated that the calculation will be very like the ordinary (scalar) case and the
reason is brought out in the next section which outlines some very basic spin--half spectral
facts. I pass on to define a calculational Dirac GJMS operator and then to evaluate its
conformal anomaly and functional determinant. The arising multiplicative anomalies are
found. An explicit formula is obtained for odd dimensions and a holographic interpretation
made. I then present some numerics based on the derived formulae. Finally, for odd
dimensions, an alternative approach, based on a Bessel Laplace transform, is given. It
yields the GJMS logdet as a quadrature. The calculation throws up an already computed
integral and gives the result as the traditional combination of Dirichlet $\eta$--functions and
$\log2$, which agrees with existing expressions.
\section{\bf 2. Dirac spectrum on spheres}
I set up the situation by recalling the essentials of the Dirac spectrum of $\nsl$ on $S^d$.
This is usually expressed in the form of eigenlevels,
  $$
   \mu_n(a)=\pm(n+a)\,,\quad n=0,1,2,\ldots,\infty\,,\quad a={d\over2}\,,
   \eql{eigs}
  $$
and degeneracies,
  $$
    g_d(n)=2^{[d/2]}\comb {n+d-1}{n}\,.
    \eql{degen}
  $$

I have introduced the parameter $a$ because it is instructive to compare these quantities,
not with those for a scalar on a {\it full} sphere but rather on a {\it hemi}--sphere.

I am interested in the (positive) iterated, or squared, Dirac operator, $\nsl^{\,2}$, with
eigenlevels $\la_n(a)=\mu_n^2(a)$ and degeneracies, $2g_d(n)$. For comparison the
eigenlevels of the scalar operator, $B^2\equiv -\De_2+(d-1)^2/4$, are
$\la_n\big((d+1)/2\big)$ for Dirichlet (D) conditions on the hemisphere rim and
$\la_n\big((d-1)/2\big)$ for Neumann (N). The degeneracies are exactly as in
(\peq{degen}), except for the spin factors.

It can thus be seen that the Dirac case is halfway between the scalar D and N cases.  Both
types of mixed boundary conditions for the Dirac field on the hemisphere yield the {\it
same} set of modes. For Dirac, adding the two hemisphere values amounts to just a factor
of two. The two boundary conditions might be termed `self--dual'.

For many purposes it was advantageous  in the scalar case not to work with the eigenlevels
and their binomial degeneracies but to write the eigenvalues in the way they emerge from
separation of variables in conjunction with the recursive, nested structure of the spherical
geometry. This yields the {\it non--degenerate} form,\footnote{ One might refer to this as
a harmonic oscillator representation.}
  $$
  \la({\bf m},a)=(a+{\bf m.}\bom)^2\,,\quad {\bom}=(1,1,\ldots,1)\,,
  \eql{lam}
  $$
so that the degeneracies of the eigenlevels $\la_n$ arise through coincidences, \ie\
$g_n=\{\sharp \,m_i:\sum_i m_i=n\}$ which gives the binomial coefficient. The same
applies to the Dirac case, which is encompassed in (\peq{lam}) if $a=d/2$, apart from the
spin factors.

The technical object used in the computation of the functional determinant is the spectral \zf\
of the propagation operator. For example, for $\nsl\,^2$ on the full sphere, using
(\peq{lam}), it is a Barnes \zf,
  $$
  \ze_{S^d}(s)=\caS\,\ze_d\big(2s,a\mid{\bf1}\big)\,,\quad a=d/2\,,
  \eql{spbarnes}
  $$
putting in the spin factors, $\caS=2^{[d/2]+1}$, by hand.\footnote{ They could be formally
included by sums over spin indices in the definition of the \zf.}

All this is quite standard, apart from the representation (\peq{spbarnes}), and I now turn to
the AdS/CFT aspect. I deal primarily with the boundary side.

\section{\bf 3. Boundary facts and Dirac GJMS operator}

Considering the round S$^d$ as a conformal boundary, the spectrum of the spin--half
two--point function can be written in terms of the usual Dirac S$^d$ quantities, (\peq{eigs})
and (\peq{degen}) as,
  $$
  \big|\La_n\big|={\Ga\big(|\mu_n(a)|-k+1/2\big)\over\Ga\big(|\mu_n(a)|+k+1/2\big)}\,,
  \eql{deigs}
  $$
for the eigenlevels, with degeneracy effectively $2g_d(n)$ . As before, $a=d/2$ and $k$ can
be considered to be a real variable subject to $0\le k<(d+1)/2$.

This has been given by Allais, [\pref{Allais}], used by Klebanov, Pufu and Safdi,
[\pref{KPS2}] and quoted in Aros and Diaz, [\pref{AaandD}]. Perhaps is not so surprising in
view of the fact that the spin--half two--point function differs from the spin--zero one by a
gamma matrix factor, and the eigenvalues vary just in the value of the parameter $a$.

This result suggests that one introduces, pragmatically, a Dirac GJMS operator (on a sphere)
by,
   $$
   D_{2k}(d)\equiv{\Ga\big(B+k+1/2\big)\over\Ga\big(B-k+1/2\big)}\,,
   \eql{dbjms}
   $$
where $B$ is the pseudo--operator,
  $$
    B= (\nsl^{\,2})^{1/2}=|\nsl|\,,
    \eql{dro}
  $$
expressed formally. The positive/negative modes of $\nsl$ will be positive/positive modes of
$|\nsl|$ for any mode expansion in spinor hyperspherical harmonics,
[\pref{DandP2,Dow30}].

An appropriate choice of $k$ for the spinor case is a {\it half integer}, $k=l+1/2$, because
$D_{2k}$ takes a product form (odd in $B$),
   $$\eqalign{
  D_{2l+1}(d)\equiv C_l\equiv
  &B\prod_{h=1}^{l}\big(B^2-h^2\big)\,\cr
  =&\prod_{h=-l}^{l}\big(B+h\big)\,,\cr
  =&B^{[2l+2]}
  }
  \eql{dbgjm2}
  $$
in terms of an (even) central factorial. Then  $D_1=C_0=B$ will give the usual (massless)
Dirac numbers. (An empty product is defined as equalling unity.)

In the scalar case, $C_l$ is a well known boundary Dirichlet--Robin (pseudo)--operator.

\section{\bf4. Zeta function and conformal anomaly}

Corresponding to (\peq{lam}), the eigenvalues of $C_{l}$ have the  form,
  $$
  \La_l({\bf m},a)=(a+{\bf m.}\bom)
  \prod_{h=1}^{l}\big((a+{\bf m.}\bom)^2-\al_h^2\big)\,,\quad
  \al_h=h\,,
  \eql{deigs2}
  $$
degenerate  only up to spin factors. Apart from the specific value of $a$, these are very
similar to the scalar eigenvalues on a D or N hemisphere, [\pref{ChandD,Dowcmp}], .
Therefore, in order to calculate the conformal anomaly and functional determinant, exactly
the same route as in [\pref{DowGJMS}] can be followed, only the value of $a$ and the
meaning of $\al_h$ have to be changed with an allowance made for the fact that one factor
is linear.

One might hope that any interpolation between the integers that arises in the course of the
calculation, would agree with the result of a computation from $D_{2k}$, (\peq{dbjms}),
but agreement {\it is} expected in odd dimensions where there is no conformal nor
multiplicative anomaly, as will be shown.

The relevant \zf\ is very similar to that defined in [\pref{DowGJMS}]
   $$\eqalign{
    Z_d(s,a,l)&=\caS\sum_{\bf m=0}^\infty{1\over (\La_{l}({\bf m},a))^s}\cr
    &=\caS\sum_{\bf m=0}^\infty{1\over(a+{\bf m.}\bom)^s}\prod_{h=1}^{l}{1\over
  \big(a+{\bf m\,.\,\bom})^2-\al_h^2\big)^s}\,,
    }
    \eql{bzeta}
  $$
and the various evaluations go through more or less {\it verbatim}. For example the value
at $s=0$ can be read off as the average,
     $$\eqalign{
  Z_d(0,a,l)&=\caS{1\over (2l+1)}\bigg(\ze_d(0,a)+\sum_{h=1}^{l} \big(
  \ze_d(0,a+\al_h)+\ze_d(0,a-\al_h)\big)\bigg)\,\cr
  &=\caS{(-1)^d\over (2l+1)\,d!}\,\bigg(B^{(d)}_d\big(d/2\big)+\sum_{h=1}^{l} \left(
  B^{(d)}_d\big(d/2+h\big)+B^{(d)}_d\big(d/2-h\big)\right)\bigg)\cr
  &=\caS{(1+(-1)^d)\,\over 2(2l+1)\,d!}\,\bigg(B^{(d)}_d\big(d/2)
  +2\sum_{h=1}^{l}B^{(d)}_d\big(d/2+h\big)\bigg)\,,
  }
  \eql{zezero}
  $$
where I have inserted the appropriate Dirac value, $a=d/2$ and used a symmetry property
of the Bernoulli polynomial. This quantity, which I will refer to as (minus) the conformal
anomaly, is thus zero in odd dimensions, as expected. Calculation of the Bernoulli numbers
quickly yields the standard Dirac conformal anomaly when $l=0$ as a check.

As in the scalar case, in this approach there is no need to derive and then expand the
degeneracies, which is the traditional method. See Cappelli and d'Appollonio,
[\pref{CandA}], and Copeland and Toms, [\pref{CandT2}], for the ordinary Dirac case.

In Fig.1 I plot the spin--half `conformal anomaly', $c_d(l)\equiv -Z(0,d/2,l)$, against $l$ for
a few low dimensions. It computes to a $d$th degree polynomial in $l$ and I have extended
the plot accordingly. The existence of negative modes does not affect this calculation. The
polynomials are even in the variable $k=l+1/2$ and I list some below.

 \epsfxsize=5truein\epsfbox{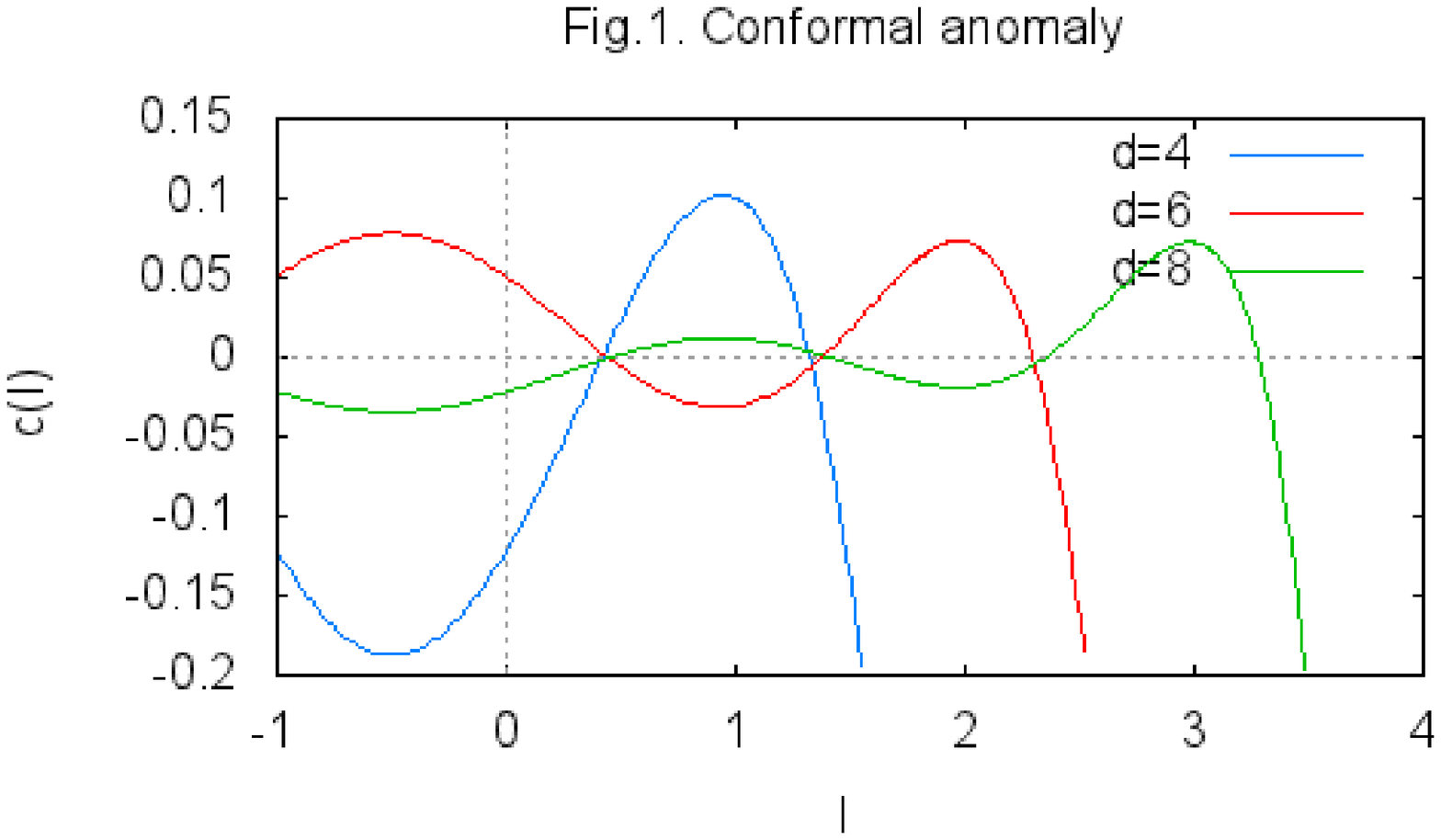}
 
  $$\eqalign{
  &c_2(k-1/2)=-{1\over6}\,(4k^2-3)\cr
 &c_4(k-1/2)=-{1\over720}\,(48k^4-200k^2+135)\cr
 &c_6(k-1/2)=-{1\over60480}\,(192k^6-2352k^4+7252k^2-4725)\cr
 &c_8(k-1/2)=-{1\over14515200}\,(1280k^8-34560k^6
 +284256k^4-774960k^2+496125)\,.
 }
 \eql{can}
  $$

I note that the curves  intersect approximately at their roots, a fact I cannot interpret.

As in [\pref{DowGJMS}] the conformal anomaly can be written more explicitly as an integral
on using specific expressions for the Bernoulli polynomials in (\peq{zezero}). I find,
  $$
  Z_d(0,a,l)+\de_{k,(d+1)/2}=\caS{2\over k\,d!}\int_0^{k}dt\,
  \prod_{i=0}^{d/2-1}\big(t^2-(i+1/2)^2\big)\,,
  \eql{cad}
  $$
where $k=l+1/2$. A holographic discussion is given by Aros and Diaz, [\pref{AaandD}].

For comparison, the spin--zero expression is,
  $$
  Z_d(0,a,l)+
  \de_{k,2d}=\caS_0{2\over k\,d!}\int_0^{k}dt\,\prod_{i=0}^{d/2-1}\big(t^2-i^2\big)\,,
  \eql{cas}
  $$
where $\caS_0=1$. In these expressions for the conformal anomaly, account has been
taken of the possible existence of zero modes.
  
\section{\bf 5. Determinants}

For zeta regularised determinants, one requires the derivative of the \zf\ at $s=0$. As
mentioned, the calculational process is virtually identical to that in [\pref{DowGJMS}] and,
therefore, so is the (intermediate) answer which is,
  $$
  Z_d'(0,a,l)=\caS\log\bigg({1\over\rho_d^{l+1}}\,\Ga_d(a)
  \prod_{h=1}^{l}\Ga_d(a+\al_h)\,\Ga_d(a-\al_h)\bigg)+\caS M^A(d,a,l)\,,
  \eql{deriv2}
  $$
where $M^A(d,a,l)$ is the multiplicative anomaly. The derivatives of the Barnes function
have been formally expressed in terms of multiple gamma functions, the definition being
  $$
  \ze_d'(0,a)=\log{\Ga_d(a)\over \rho_d}=\log G_d(a)\,.
  \eql{defgam}
  $$
$\rho_d$ is a normalising modulus, [\pref{Barnesa}], and is independent of $a$.

For the moment, I put the multiplicative anomaly to one side and compactify the product in
(\peq{deriv2}) by using the recursion,
$$
  {\Ga_d(a)\over\rho_d}={\Ga_{d+1}(a)\over\Ga_{d+1}(a+1)}\,,
  \eql{reln}
  $$
which has the effect of bringing in the next higher dimension.
 
Because I gave no details in [\pref{DowGJMS}] I take the opportunity here for a brief
explanation.

Substituting (\peq{reln}) into (\peq{deriv2}) yields,
  $$
  Z_d'(0,a,l)=\caS\log\bigg({\Ga_{d+1}(a)\over\Ga_{d+1}(a+1)}
  \prod_{h=1}^{l}{\Ga_{d+1}(a+h)\,\Ga_{d+1}(a-h)
  \over\Ga_{d+1}(a+h+1)\,\Ga_{d+1}(a-h+1)}\bigg)+\caS M^A\,,
  \eql{deriv4}
  $$
and it is clear there will be considerable cancellation. I find the simple expression, with a
holographic flavour,
  $$
  Z_d'(0,a,l)=\caS\log{\Ga_{d+1}(a-l)\over\Ga_{d+1}(a+l+1)}+\caS M^A(d,a,l)\,.
  \eql{deriv5}
  $$
  
This holds whatever the value of $a$. In particular, $a=d/2$ and $l=0$ gives the ordinary
Dirac case, (obtainable directly of course). There can be no multiplicative anomaly, reflected
in $M^A(d,d/2,0)=0$. Aros and Diaz, [\pref{AaandD}], give this logdet in the form,
  $$
  Z_d'(0,a,0)=\caS\log G_{d}(a)\,,\quad a=d/2\,,
  $$
which follows using (\peq{reln}). Klebanov, Pufu and Safdi [\pref{KPS2}] have given the
results of calculations of the determinant based on (\peq{deigs}), up to a factor, using the
dimensional regularisation approach of Diaz and Dorn, [\pref{DandD}].
  
 \section{\bf 6. The other factorisation}
 
In AdS/CFT the problems that one wishes to solve are associated with the \zf\ of the `exact'
eigenvalues, (\peq{deigs}), \ie,
  $$
   \Xi_d(s,a,l)\equiv\caS\sum_{\bf m=0}^\infty{1\over (\La(l,{\bf m},a))^s}\,,
   \eql{xi}
  $$
where, in non--degenerate form,
  $$
   \La(l,{\bf m},a)={\Ga\big(a+{\bf m.}\bom+l\big)
   \over\Ga\big(a+{\bf m.}\bom-l+1\big)}\,.
   \eql{lamb}
  $$
Here $l$ is now a real variable subject to $1\le2l<d+2$. (For the Dirac case, $a=d/2$).
  
Whenever $\La$ can be written as a product, which happens when $l$ is either integral or
half--integral, all calculations can be effected explicitly, as above. The latter case
corresponds to the GJMS operator $P_{2k}$ treated in [\pref{DowGJMS}], but applies here
also. It gives the alternative product form (now even in $B$),
  $$\eqalign{
  \ol C_{k}(d)
  =\prod_{h=0}^{k-1}\big(B^2-\al_h^2\big)\,,\quad \al_h=h+1/2\,,\cr
  }
  \eql{prod2}
  $$
instead of (\peq{dbgjm2}). The non--degenerate eigenvalues are
    $$
  \ol\La_k({\bf m},a)=
  \prod_{h=0}^{k-1}\big((a+{\bf m.}\bom)^2-\al_h^2\big)\,,
  \eql{deigs3}
  $$
and the \zf\ is, of course, a special case of (\peq{xi}),
  $$
   \ol Z_d(s,a,k)=\caS\sum_{\bf m=0}^\infty{1\over (\ol\La_{k}({\bf m},a))^s}\,.
  $$
  
The technical situation is now identical to that in [\pref{DowGJMS}] and yields the
intermediate formula,
  $$
  \ol Z_d'(0,a,k)=\caS\log
  \prod_{h=0}^{k-1}{\Ga_{d+1}(a+h+1/2)\,\Ga_{d+1}(a-h-1/2)
  \over\Ga_{d+1}(a+h+3/2)\,\Ga_{d+1}(a-h+1/2)}+\caS M^B\,.
  \eql{deriv6}
  $$
Simplification, using (\peq{reln}), results in,
   $$
 \ol Z_d'(0,a,k)=\caS\log{\Ga_{d+1}(a-k+1/2)\over\Ga_{d+1}(a+k+1/2)}
 +\caS M^B(d,a,k)\,,
  \eql{deriv7}
  $$
and the point now is that the non--polynomial part of this is {\it identical} to that in
(\peq{deriv5}) showing that, apart from any multiplicative anomaly, the interpolations of
$\Xi'_d(0,l)$ provided by the two factorisations are the same. I am not certain of the
spectral significance of the structure, (\peq{prod2}).
 \section{\bf 7. The multiplicative anomaly}
 
I now turn to the multiplicative anomaly correction, $M^A(d,a,k)$, a  means of determining
which, for the product form (\peq{prod2}), has been given in [\pref{DowGJMS}] equn.
(16). Unfortunately an algebraic error means that the expression for $M_2$ there is
incorrect. The correct version can be found in [\pref{DowGJMSE}] and will not be shown
here. However I give the corresponding formulae for the product (\peq{dbgjm2}) obtained
from (\peq{bzeta}). The answer is very similar. The extra single factor in (\peq{deigs2}) is
easily allowed for and I find,
  $$
  M^B(d,a,l)=M^B_1(d,a,l)+M^B_2(d,a,l)\,,
  $$
where
  $$
  M^B_1(d,a,l)=-\sum_{r=1}^u{1\over r}\bigg(\sum_{h=1}^{l}\al_h^{2r}\bigg)
  H_{l+1/2}(r)\,N_{2r}(d,a)\,,
  \eql{mb1}
  $$
and
  $$
  M^B_2(d,a,l)={1\over2l+1}\sum_{r=1}^u{1\over r}\sum_{t=1}^{u-r}
  {1\over t}\bigg(\sum_{i<h=0}^{l-1}
  \al_i^{2r}\,\al_h^{2t}\bigg)\,N_{2r+2t}(d,a)\,.
  \eql{mb2}
  $$
The upper limit $u$ equals $d/2$ for even dimensions, and $(d-1)/2$ for odd, \ie\ $[d/2]$.

$H_k$ is related to the harmonic series, $H(r)=\sum_{n=1}^r1/n$ by,
  $$
  H_k(r)=H(2r-1)-{1\over2k}H(r-1)\,,
  $$
and $N$ is the residue at the pole of the Barnes \zf,
    $$
  N_r(d,a)={1\over(r-1)!(d-r)!}B^{(d)}_{d-r}(a)\,.
  $$
 
These formulae hold for any $a$ and any distribution of the $\al_h$. To repeat, for $M^A$,
$\al_h=h+1/2$ while for $M^B$, $\al_h=h$. For these values, the sums over the $\al$s in
(\peq{mb1}) and (\peq{mb2}) have a combinatorial significance which I will not invoke at
this time.

The sums for $M^A$ and $M^B$ can be performed for a given $d$, and $a=d/2$ for Dirac,
to give polynomials in $k$ and $l$ respectively. Explicit calculation reveals the equality,
  $$
  M^A(d,d/2,k)=M^B(d,d/2,l)\,,
  $$
if $k=l+1/2$, even though $k$ and $l$ have to be integers  for the intermediate
summations to make sense.

Looking back to (\peq{deriv5}) and (\peq{deriv7}) confirms that the full effective actions
for the two factorisations are also related algebraically by,
  $$
  Z_d'(0,a,l)= \ol Z_d'(0,a,k)\,,\quad{\rm if}\quad k=l+1/2\,.
  $$
This gives us confidence to extend the results from the integers to the reals to give the
quantity $\Xi'(0,a,l)$, as a computable quantity, say (\peq{deriv5}),
  $$
  \Xi'(0,a,l)=Z'(0,a,l)\,.
  $$
  
In odd dimensions, the multiplicative anomaly vanishes by a symmetry property of the
Bernoulli polynomials, just like the conformal anomaly.

In even dimensions, for completeness, I exhibit a few of the polynomials, as functions of
$k$, with a factor of $k (1-4k^2)$ removed,
  $$\eqalign{
  M^A(2,1,k)&={1\over12}\cr
  M^A(4,2,k)&={1\over2160}\,(14k^2-39)\cr
  M^A(6,3,k)&={1\over7257600}\,(1392k^4-14016k^2+28745)\cr
  M^A(8,4,k)&={1\over1524096000}\,(4656k^6-111096k^4+749695k^2-1355760)\,.
    }
   $$
 
 A polynomial  ambiguity occurs in other schemes for computing $\log\det D_{2k}$ and
$\log\det P_{2k}$, [\pref{DandD,Diaz}]. It is tolerably clear that the multiplicative
anomalies displayed here and in [\pref{DowGJMSE}] are unique, within the present scheme.
Given that they should be odd in $k$ and zero for $k=1/2$ means that what remains is a
polynomial of degree $d/2-1$ in $k^2$ and the unknown coefficients can be determined,
with checks, from the values of the polynomial at integer, or at half--integer, $k$, which are
known with certainty.

The physical significance of any multiplicative anomaly is problematic because of the
uncertainties raised by the necessity of renormalisation.\footnote{ A similar comment
applies in the AdS/CFT context where eigenvalue prefactors are routinely argued away.} It
is also a quantity manufactured on the \zf\ definition of the functional determinant. Different
definitions can give different continuations. However if the \zf\ method is used consistently
to compare various quantities then the correct, \ie\ appropriate, multiplicative anomalies
must be employed.

Because, in \zf\ regularisation, the conformal anomaly drives the infinities (and/or log
term), there might be a case for endowing the multiplicative anomalies at the roots of
$c_d(l)$, (\peq{can}), with some significance. As displayed in Fig.1 these roots appear to be
remarkably stable with varying dimension.

\section{\bf 8. Odd dimensions}
Here there is no multiplicative anomaly and the effective action can be given in terms of
elementary functions. I follow exactly the procedure of [\pref{DowGJMS}].

The expression (\peq{deriv5}), equivalently (\peq{deriv7}), is written in terms of
Kurokawa's  multiple sine function, essentially by definition,\footnote{ I refer to
[\pref{DowGJMS}] for information on this, and other, multiple functions, with original
references.}
  $$\eqalign{
   Z_d'(0,a,l)&=\caS\log\sin_{d+1}(a+k+1/2)\cr
   &={1\over2}\caS\big(\log\sin_{d+1}(a+k+1/2)
   -\log\sin_{d+1}(a-k+1/2)\big)\cr
   &={1\over2}\caS\int_{(d+1)/2-k}^{(d+1)/2+k}dz\,\cot_{d+1}(z)\cr
   &=-{1\over2d!}\caS\int_{(d+1)/2-k}^{(d+1)/2+k}dz\,B^{(d+1)}_d(z)
   \,\pi\,\cot(\pi z)\cr
   &={\caS\over d!}\int_{0}^{k}dz\,B^{(d+1)}_d\big((d+1)/2+z\big)\, \pi\cot(\pi z)\cr
   &\equiv\caS\int_{0}^{k}dz\,P_s(d,z) \pi\cot(\pi z)\,,\cr
   }
  \eql{deriv8}
  $$
where the polynomial, $P_s$, is  defined by,
  $$
  P_s(d,z)={1\over d!}\,z\prod_{i=1}^{(d-1)/2}(z^2-i^2)\,.
  \eql{plpoly}
  $$
  
I return to this expression later. It can be used for numerical computation but I prefer the
representation in terms of the digamma function that works for both odd and even
dimensions. I gave some numbers using this for the spin zero case in [\pref{DowGJMSE}]
and section 11 below contains a few Dirac values.

For the three--sphere, (\peq{deriv8}) agrees with the expression given by Klebanov, Pufu
and Safdi, [\pref{KPS2}], equn.(66) derived using Diaz and Dorn, [\pref{DandD}]. Their
$\De$ equals $3/2-k$ here. The intermediate details would appear to be less elegant.

\section{\bf 9. Holographic interpretation}

Purely by algebra we have reached an equation, (\peq{deriv7}), which has a holographic
look in that the right--hand side refers to one higher dimension. To bring this out explicitly,
show the dimension of the operator (\peq{dro}) by $B_d$ and note that the \zf\ of
$B_{d+1}+k$ is, \cf\ (\peq{spbarnes}), $2\ze_d(s,d/2+k+1/2)$ so that,
  $$\eqalign{
  \log{\det B_{d+1}+k\over\det B_{d+1}-k}&=\caS_{d+1}\log{\Ga_{d+1}(d/2+k+1/2)
  \over\Ga_{d+1}(d/2-k+1/2)}\cr
  &=-2Z_d'(0,d/2,k)+\caS_{d+1} M(d,d/2,k)\,,
  }
  $$
using (\peq{deriv7}) and $\caS_{d+1}/\caS_d=2$. The left--hand side refers to operators
on the $(d+1)$--sphere.

In odd dimensions there results the $(d+1)$ identity,
  $$
  \log{\det B_{d+1}+k\over\det B_{d+1}-k}
  =\caS_{d+1}\int_{0}^{k}dz\,P_s(d,z) \pi\cot(\pi z)\,,
  $$
and, as pointed out in [\pref{DowGJMS}] in the scalar  case, by differentiating with respect
to $k$ it follows that the residues (times $\caS_{d+1}$) at the poles of the integrand, which
are at $(d+1)/2+n$, $n=0,1,2\ldots$, are the degeneracies of the eigenlevels
$\mu_n(a)=a+n$ of $B_{d+1}$, ($a=(d+1)/2$). By a general theorem, this implies that
the integrand is proportional to the spin--half Plancherel measure on the space dual to
S$^{d+1}$, \ie\ on the hyperbolic H$^{d+1}$, and this can be checked from the calculated
form, (\peq{plpoly}). In fact this could be taken as a {\it derivation} of the Plancherel
measure.

\section{\bf 10. Some formal remarks}

In the, more discussed, scalar case, in free field CFT the two point function on the sphere,
S$^d$, is given as (\eg\ Gubser and Klebanov, [\pref{GandK}]),
  $$
  \av {\caO(\xi)\caO(\eta)}={1\over |\xi-\eta|^{d-2k}}\,,
  \eql{tpf}
  $$
where $\xi$ and $\eta$ are the unit vectors of two points on the sphere and $|\xi-\eta|^2=
2(1-\cos\th)$ in terms of the radial angle, $\th$, between these points.

The expression (\peq{tpf}) can be derived by conformal transformation from flat space
where one has the standard Riesz potential,
  $$
\big(  \De_2^{-k} f\big)(x)={1\over\ga(k)}\int_{{\bf R}^d}|x-y|^{2k-d}f(y)\,dy\,,
\eql{rp}
  $$
with normalisation,
  $$
  \ga(k)=2^{2k}\pi^{d/2}{\Ga(k)\over\Ga(d/2-k)}\,.
  $$

Now the stereographic conformal transform of $\De_2^k$ is the GJMS operator, $P_{2k}$,
on the sphere as detailed explicitly by Graham, [\pref{Graham}]. So, taking the conformal
transform of (\peq{rp}) \ie,
  $$
\big(  P^{-1}_{2k} f\big)(\xi)={1\over\ga(k)}\int_{S^d}|\xi-\eta|^{2k-d}f(\eta)\,d\eta\,,
\eql{rp2}
  $$
would solve the eigenproblem immediately, given $P_{2k}$ as the ratio of
$\Ga$--functions, as in (\peq{dbjms}), if this were known independently, say by Branson's
intertwiner group theory method.

However, if one wishes to proceed stereographically, it is necessary to note that Graham
obtains only the (original) {\it product} form of $P_{2k}$, as in (\peq{prod2}). In order to
produce the, more general,ratio of $\Ga$--functions, with $B$ related to the
Yamabe--Penrose operator, one can apply the Funk--Hecke theorem to the kernel in
(\peq{rp2}). This was done by Morpurgo, [\pref{Morpurgo}], using an existing Gegenbauer
expansion, [\pref{EMOT2}] \S10.20 (6) (he actually quotes equn.(12)), to find the operator
$\Ga$--function form of $P_{2k}$ from (\peq{tpf}) by showing that the eigenlevels have
the structure (\peq{deigs}), for suitable $\mu_n$ with the usual sphere degeneracies ($k$
will be reversed). The same calculation was done by Gubser and Klebanov, [\pref{GandK}]
and Allais, [\pref{Allais}], for spin--half, in the physical CFT context.
\section{\bf 11. A few computations}

I now give some numerical results by the same method as in [\pref{DowGJMSE}]. Despite
the unknown physical status of the multiplicative anomaly, I present values of the derivative
of the \zf\ at zero, say from (\peq{deriv7}), and still call this the effective action, for short.

I repeat the basic equations,
  $$\eqalign{
  \log{\Ga_d(z_2)\over\Ga_d(z_1)}&=\int_{z_1}^{z_2}dz\,\psi_d(z)\,,\cr
  }
  \eql{lograt}
  $$
 with,
   $$
    \psi_d(z)={(-1)^{d-1}\over(d-1)!}
  \bigg(B_{d-1}^{(d)}(z)\,\psi(z)+Q_d(z)\bigg)\,,
  \eql{psid}
   $$
 where the polynomial $Q$ is given by,
  $$
  Q_d(z)=-(-1)^{d-1}
  \sum_{n=1}^{d-1}{(-1)^n\over n}\,B_{d-n-1}^{(d-n)}(d-z)\,B_{n}^{(n)}(z)\,,
  $$
in terms of generalised Bernoulli polynomials. For (\peq{deriv7}), $z_1=(d+1)/2-k$,
$z_2=(d+1)/2+k$.

\epsfxsize=5truein \epsfbox{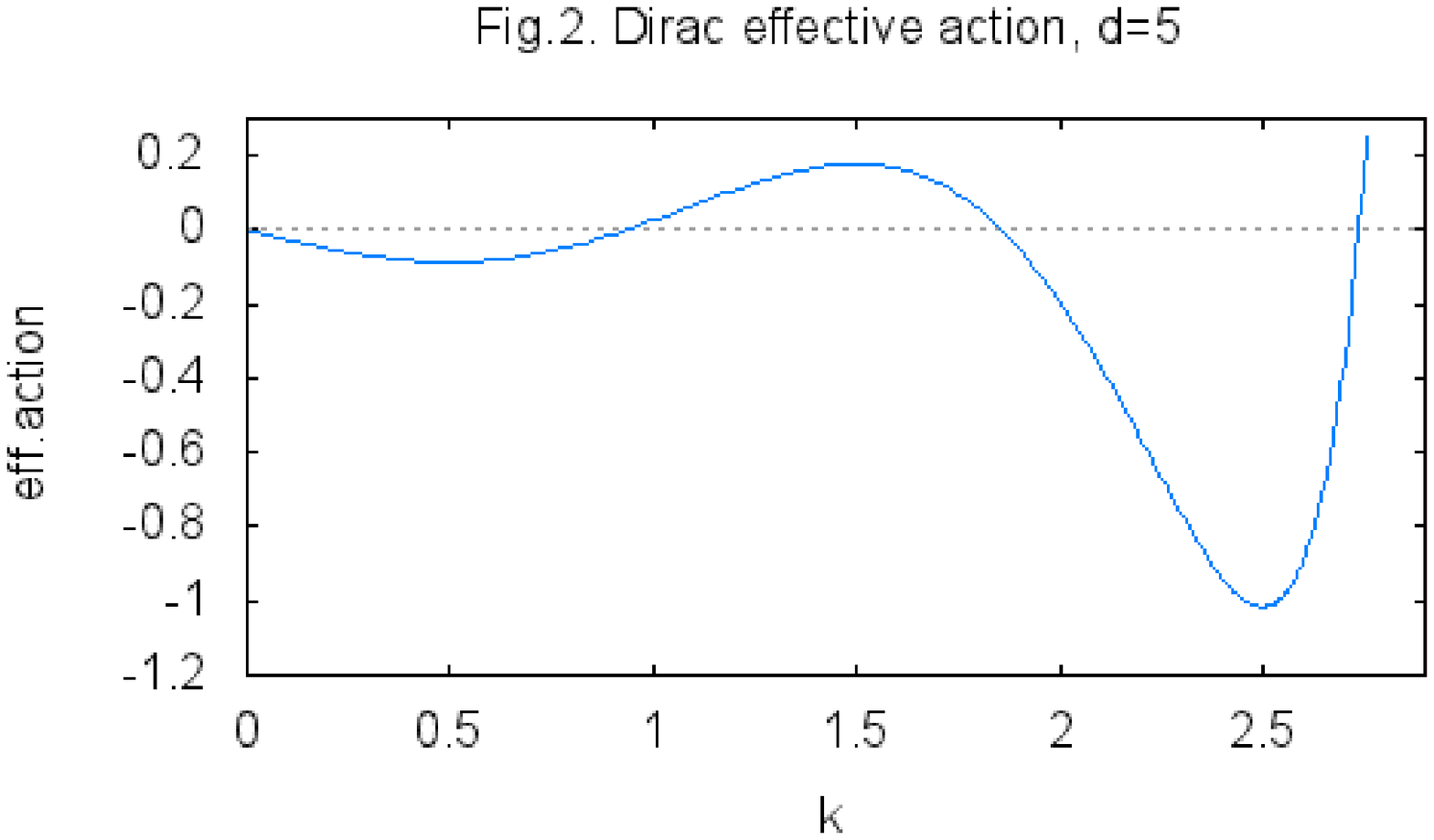}

\vglue .25in

\epsfxsize=5truein \epsfbox{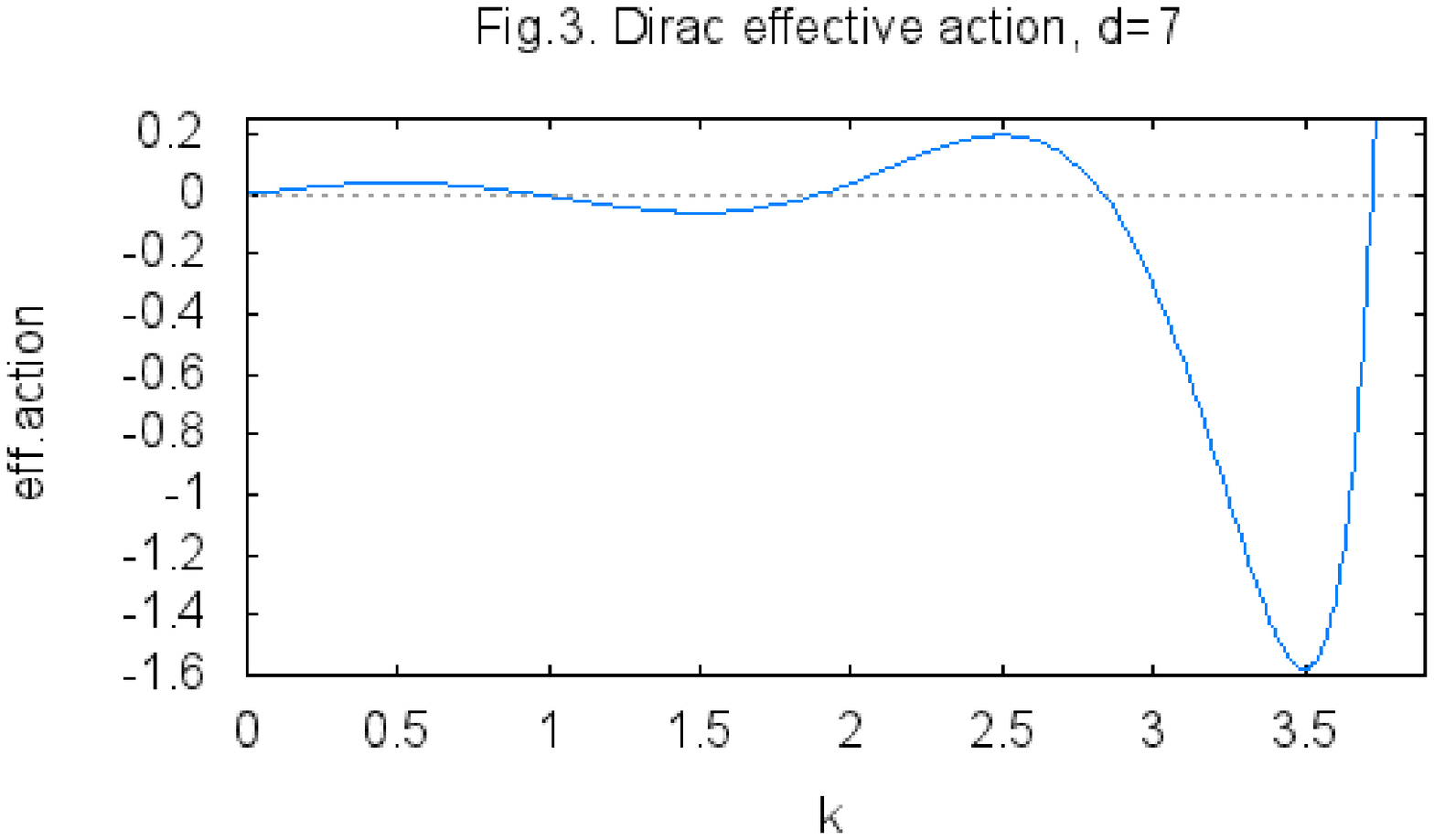}

\vglue .25in

Figures 2 to 4 plot the effective action for several dimensions over the relevant ranges of
$k$ and show a series of extrema at half integers. I recall that $k=1/2$ corresponds to the
usual Dirac case. The values agree with those that are available in [\pref{KPS2}]. Figure 5
is for an even dimension ($d=6$) the two curves illustrating the effect of leaving out the
multiplicative anomaly.

\epsfxsize=5truein \epsfbox{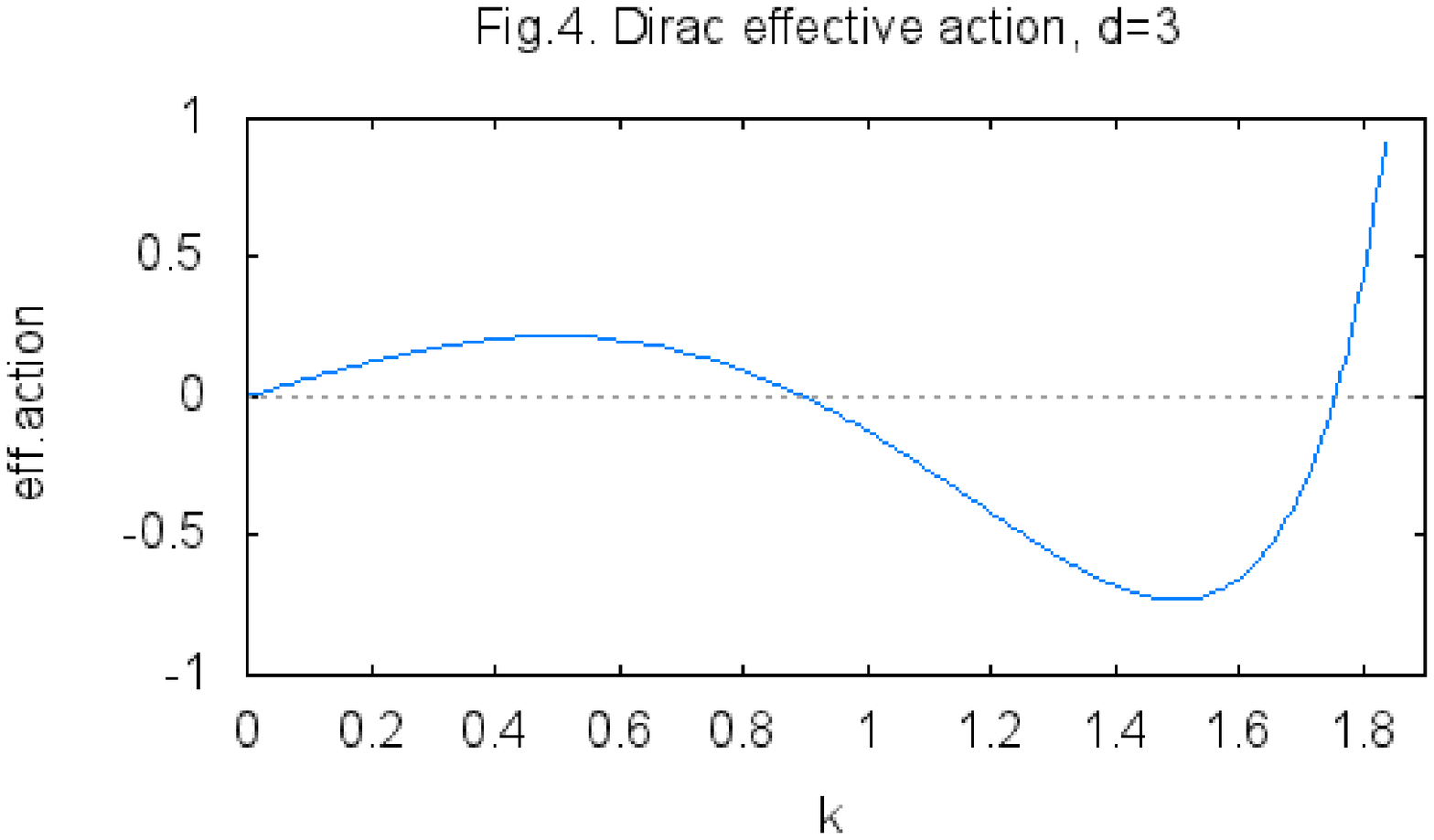}

\vglue .25in

\epsfxsize=5truein \epsfbox{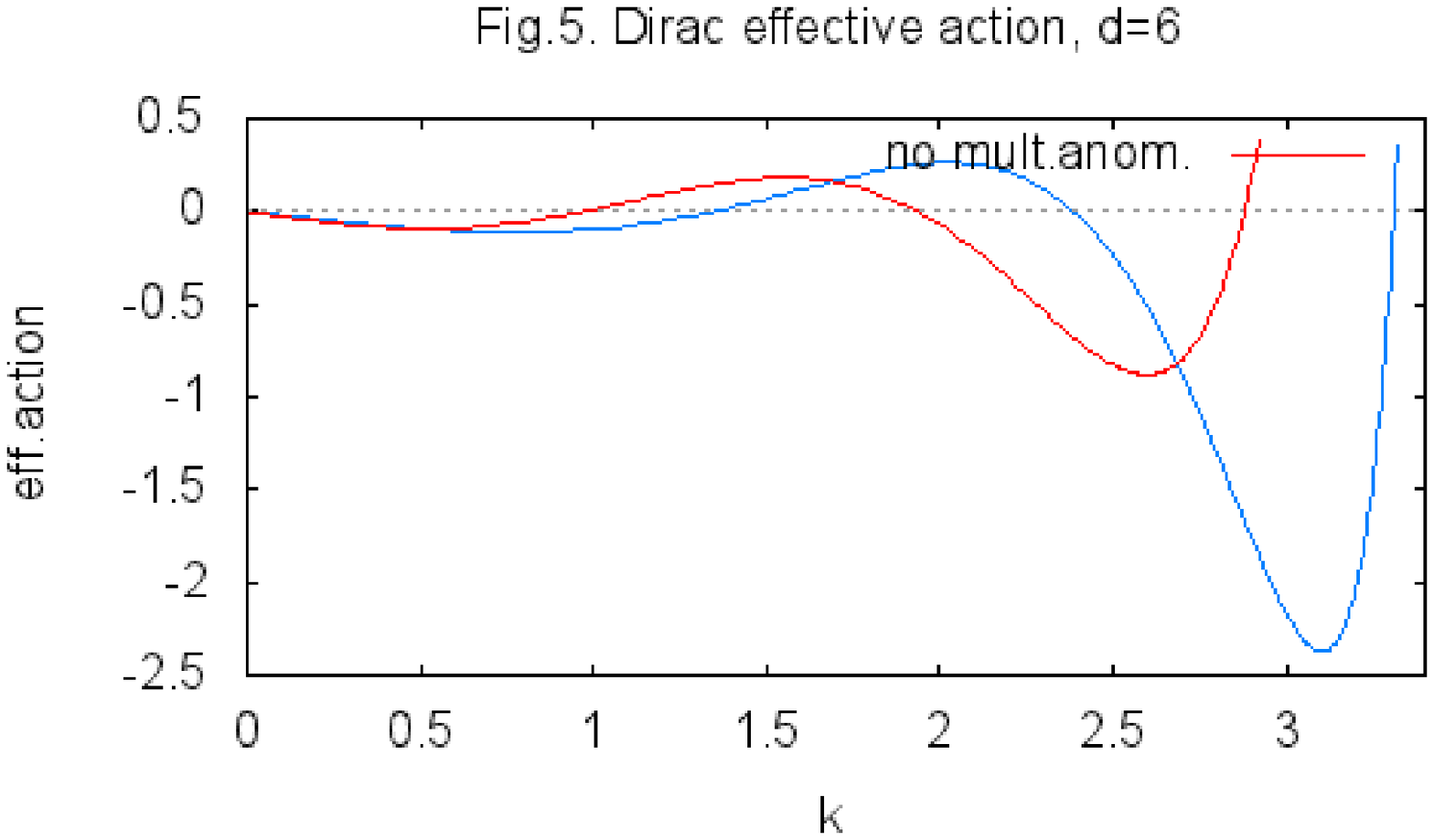}

\section{\bf 12. Alternative treatment of determinant for odd dimensions}

In [\pref{Dowren}] I gave an expression for the scalar determinant on odd spheres in the
form of an integral, different from the one corresponding to (\peq{deriv8}), which was also
suitable for numerical calculation. In this section I wish to do the same for the Dirac GJMS
case. The details are a little different as I deal, first, with the factorisation (\peq{dbgjm2})
so that the ordinary Dirac values can be recovered when $l=0$.

From (\peq{dbgjm2}),
  $$
   Z_d'(0,a,l)=-\log\det C_l=-{1\over2}\log\det B^2-\sum_{h=1}^l \log\det (B^2-h^2)\,,
   \eql{logdetC}
  $$
using, again, the absence of a  multiplicative anomaly in odd dimensions.

The expression is treated using a Bessel Fourier transform as in [\pref{Dowren}] where
proper references can be found. The detailed algebra is in [\pref{Dowren}] and so I just
give the result I want,
  $$
  -\log\det (B^2-h^2)=\caS_d{1\over2^{d-1}}\int_{\caC}d\tau\,{
  \cosh(\om-a)\tau\,\cosh h\tau\over\tau\,\prod_{i=1}^d\sinh(\om_i\tau/2)}\,,
  $$
where the contour, $\caC$, runs above all the real axis and below the first singularity of the
integrand with a positive imaginary part.
  
For the total expression, (\peq{logdetC}), the sum  gives the Dirichlet kernel,
  $$
  -\log\det C_l=\caS_d{1\over2^{d}}\int_\caC d\tau\,{
  \cosh(\om-a)\tau\,\over\tau\,\prod_{i=1}^d\sinh(\om_i\tau/2)}\,U_{2l}(\cosh\tau/2)\,,
  $$
where $U_{2l}$ is a Chebyshev polynomial. Referring to the eigenvalues (\peq{eigs})
$a=d/2$ and since, for the round sphere, $\om\equiv\sum_i\om_i/2=d/2$ we see that the
$\cosh$ factor gives unity. So, on the $d$--sphere, for spin--half,
  $$
   F_l(d)\equiv-{1\over\caS_d}\log\det C_l(d)={1\over2^{d}}\int_\caC d\tau\,{
  1\over\tau\,\sinh^d(\tau/2)}\,U_{2l}(\cosh\tau/2)\,,
  \eql{Fd}
  $$

For comparison I give the corresponding spin-0 result,[\pref{MandD}],
  $$
\log\det P_{2k}(d)=
-{1\over 2^{d-1}}\int_\caC d\tau\,
 {1\over\tau\,\sinh^d(\tau/2)} \cosh(\tau/2)\,U_{2k-1}\big(\cosh\tau/2\big)\,.
 \eql{sp0}
  $$

Now, as in the earlier, scalar calculation, I re--expand the Chebyshev polynomial, this time
in even powers of $\sinh\tau/2$,
  $$\eqalign{
  U_{2l}(\cosh\tau/2)&=2\sum_{i=0}^{l}{(-1)^i\,2^{2i}\over(2 i+1)!} (l+1/2)^{[2i+1]}
  \sinh^{2i}(\tau/2)\cr
  &\equiv\sum_{i=0}^{l} 2^{2i}\,A_i(l)\,\sinh^{2i}(\tau/2)\,,
  }
  $$
using central factorial notation.

Hence one obtains the decomposition sum rule,
  $$
  F_l(d)=\sum_{i=0}^l A_i(l)\,F_0(d-2i)
  \eql{sumrule}
  $$
in terms of the {\it ordinary} Dirac values, $\sim F_0$, at dimensions $\le d$ and $\ge
d-2l$. This would be one way to evaluate the Dirac GJMS determinants. Numerically it is less
efficient than (\peq{Fd}), but has algebraic consequences.

Before discussing these I return, for comparison, to the scalar case, expressed slightly
differently to [\pref{MandD}]. The expansion required is now,
  $$\eqalign{
  U_{2k-1}(\cosh\tau/2)&=2\cosh(\tau/2)
  \sum_{i=0}^{k-1}{(-1)^i 2^{2i}\over(2i+1)!}\,k^{[2i+2]-1}\,\sinh^{2i}(\tau/2)\cr
  &\equiv \cosh(\tau/2)\sum_{i=0}^{k-1} B_i(k)\,\sinh^{2i}(\tau/2)
  }
  \eql{even}
  $$

Substituted into (\peq{sp0}), this gives
  $$
  -\log\det P_{2k}={1\over 2^{d-1}}\int_\caC {d\tau\over\tau}
  \bigg({1\over\sinh^d(\tau/2)}+{1\over\sinh^{d-2}(\tau/2)}\bigg)
  \sum_{i=0}^{k-1} B_i(k)\,\sinh^{2i}(\tau/2)\,,
  $$
which shows that the basic integral required is
   $$
    \int_\caC d\tau\,{  1\over\tau\,\sinh^d(\tau/2)}
    =2^d\, F_0(d)\,,
   $$
from (\peq{Fd}) and, therefore, that the scalar GJMS determinant is also expressed as a
sum of ordinary {\it Dirac} determinants. The lowest order case is
   $$
  -\log\det P_2(d)=2\big(F_0(d)+F_0(d-2)\big)\,,
  \eql{loc}
  $$
where $P_2$ is the Yamabe--Penrose conformally invariant Laplace operator. This relation
reflects properties of the eigenvalues. It can be inverted to give the spin--1/2 quantity in
terms of a sum of scalar ones at varying dimensions,
  $$
   F_0(d)={1\over2}\sum_{i=1}^d(-1)^{i+1} \log\det P_2(i)
  $$
which is of no numerical interest but has a curiosity value.

In a similar vein I note that the basic recursion for the Chebyshev polynomials (just
trigonometry),
$$
U_{n+1}(x)=2xU_n-U_{n-1}(x)\,,
$$
gives a recursion on the {\it order} of the operators,
$$
-\log\det P_{2k}(d)=F_{k}(d)+F_{k-1}(d)\,,
$$
again somewhat of a curiosity.

As mentioned, the ordinary Dirac values are known but I will discuss them again using the
approach of this section.

So far as explicit computation is concerned one way is to transcribe the contour integral into
real form by setting $\tau=x+i\pi$ and using the symmetry of the integrand to give
   $$\eqalign{
   F_0(d)&={(-1)^{(d+1)/2}\over2^{d-1}}\int_0^\infty dx\, {\sech^d(\pi x/2)\over x^2+1
   }\cr
   &\equiv {(-1)^{(d+1)/2}\over2^{d-1}}\,J(d)\,,
   }
   \eql{F0d}
  $$
where $J(d)$ is the same integral that occurs in the scalar case, [\pref{Dowren}], in
accordance with the expression (\peq{loc}).

The integral $J(d)$ was obtained in [\pref{Dowren}] as a combination of Riemann \zfs\
(actually, Dirichlet $\eta$--functions) and $\log2$ by a method which involved expanding
the $\sech^d$ in derivatives of $\sech$. The coefficients appeared as, [\pref{Dowcen}],
central differentials of zero, $D^r\,O^{[s]}$ (\eg\ Steffensen, [\pref{Steffensen}]),
$$
\eqalign{
 J_{2m+1}=
(-1)^m\sum_{n=1}^m(-1)^{n}\,&{\frac{2^{2(m-n)}\,D^{2n+1}0^{[2m+1]}}
{\pi^{2n}(2m)!(2n+1)}}\,\eta(2n)\cr
&+(-1)^m\,D0^{[2m+1]}\,\log2.
}
  $$

A similar form results from pushing the contour $\caC$ in (\peq{Fd}) upwards to infinity,
and employing residues, in the fashion of Candelas and Weinberg, [\pref{CaandWe}]. A
systematic discussion, starting from (\peq{F0d}), is given in [\pref{MandD}] the coefficients
now appearing as higher $D$--N\"orlund (Bernoulli) numbers from the power series of
$\cosech^d$ (see below). The equality of the two expressions is guaranteed by (or is a
proof of) an identity between these numbers and the differentials of zero (or, equivalently,
the central factorial numbers of the first kind), see [\pref{MandD}].

The contour method in [\pref{MandD}] effectively employs the decompositions like
(\peq{loc}), the quantity denoted  there  by $f_{2m+1}$ equalling $J(2r+1)/\pi$ in
(\peq{F0d}).

The expressions agree with those, \eg, in [\pref{KPS2}] Table 2, derived by a more
conventional eigenvalue method.

The same Bessel technique can be applied to the other factorisation of the `intertwiner',
(\peq{prod2}). The algebra is very similar to the scalar case, except that now $a=d/2$ so
that, instead of (\peq{sp0}), there appears, ($k\in\oZ$),
  $$
-\log\det \ol C_{k}(d)={\caS_d
\over 2^{d-1}}\int_\caC d\tau\,
 {1\over\tau\,\sinh^d(\tau/2)}\, U_{2k-1}\big(\cosh\tau/2\big)\,.
 \eql{sp0}
  $$

As before, this integral can be treated numerically as it stands but again I employ the
expansion (\peq{even}) to give,
  $$\eqalign{
  -\log\det \ol C_{k}(d)&={\caS_d\over 2^{d-1}}\sum_{i=0}^{k-1}B_i(k)
  \int_\caC d\tau\,{\cosh(\tau/2)\over\tau\,\sinh^{d-2i}(\tau/2)}\cr
  &=-{\caS_d\over 2^{d-2}}\sum_{i=0}^{k-1}{B_i(k)\over d-2i-1}
  \int_\caC d\tau\,{1\over\tau^2\,\sinh^{d-2i-1}(\tau/2)}\,,
  }
  \eql{altdet}
  $$
which, out of interest, I evaluate by residues using the standard expansion,
   $$
   (t\,\cosech t)^d=\sum_{\nu=0}^\infty {D^{(d)}_{2\nu}\over(2\nu)!} \,t^{2\nu}\,,
   $$
 in terms of $D$--N\"orlund numbers. Basic algebra gives the double summation,
   $$\eqalign{
  &-{\log\det \ol C_{k}(d)\over\caS_d}={(-1)^{(d-1)/2}
  \over 2^{d-1}}\sum_{i=0}^{k-1}{(-1)^i\over\pi^{d-2i-1}}\,B_i(k)\bigg[\cr &
  \sum_{\nu=0}^{(d-3-2i)/2}{(-1)^\nu\over(2\nu)!}
  { d-2i-1-2\nu\over  d-2i-1}\,D^{(d-2i-1)}_{2\nu}\,\pi^{2\nu}\,\ze_R(d-2i-2\nu)\bigg]\,,
  }
  \eql{altdet2}
  $$
which is a sum of Riemann, not Dirichlet, zetas. Also there is no $\log2$.

The simplest case is the operator for $k=1$ \ie\ $\ol C_1(d)= \nsl^{\,2}-1/4$ from
(\peq{prod2}). For example, the right--hand side of (\peq{sp0}) reduces to
$\ze_R(3)/2\pi^2$ for $d=3$ and to $-\ze_R(3)/24\pi^2-\ze_R(5)/8\pi^4$ for $d=5$, and
so on.\mgn{Check signs} The next case is the Paneitz operator when $k=2$, \ie $\ol
C_2(d)=(\nsl^{\,2}-1/4)\,(\nsl^{\,2}-9/4)$. For $d=5$ there results
$\ze_R(5)/4\pi^4-5\ze_R(3)/12\pi^2$. Numerical quadrature from the real form of
(\peq{altdet}) gives agreement to at least 14 places. While this is really only a check of the
algebra it does reveal that quadrature is more efficient.

\section{\bf 13. Conclusion and comments}
The conformal anomalies and functional determinants have been derived, and computed, for
spin--half GJMS operators on spheres using a direct spectral method directly parallel to the
corresponding scalar case. Technically, I have again found that the Barnes \zf\ is very handy
allowing one to avoid expanding degeneracies and the resulting Riemann/Hurwitz \zfs.
However, these reappear via a different route.

The determinants calculated for two factorisations of the general $\Ga$--function form of the
GJMS operator (the intertwiner) were shown to be equal when continued, including the
multiplicative anomalies.

The results, which have a holographic aspect, can also be derived formally by a tensor
product construction as in [\pref{DowGJMS}] but I leave this for another time. I also
postpone discussion of the fact that counting the number of {\it negative} modes, as $k$
varies, of the GJMS operator on the $d$--sphere gives the degeneracies of $\nsl^2$ on the
$(d+1)$--sphere.

An alternative approach for odd dimensions yielded the answer for the (log) determinant as
a quadrature. Further analytical work turned this into the familiar sum of Dirichlet $\eta$--
functions and $\log2$. For the other factorisation, just a sum of Riemann zetas appears.

 \vglue 20truept

 \noin{\bf References.} \vskip5truept
\begin{putreferences}
   \ref{Dowcen}{Dowker,J.S., {\it Central differences, Euler numbers and symbolic methods},
 \break ArXiv:1305.0500.}
 \ref{EMOT2}{Erdelyi, A., Magnus, W., Oberhettinger, F. and Tricomi, F.G. {
  \it Higher Transcendental Functions} Vol.2 (McGraw-Hill, N.Y. 1953).}
 \ref{Graham}{Graham,C.R. SIGMA {\bf 3} (2007) 121.}
  \ref{Morpurgo}{Morpurgo,C. \dmj{114}{2002}{477}.}
      \ref{DandP2}{Dowker,J.S. and Pettengill,D.F. \jpa{7}{1974}{1527}}
 \ref{Diaz}{Diaz,D.E. {\it JHEP} {\bf 0807} (2008) 103.}
    \ref{DandD}{Diaz,D.E. and Dorn,H. {\it JHEP} {\bf 0705} (2007) 46.}
    \ref{AaandD}{Aros,R. and Diaz,D.E. {\it Determinant and Weyl anomaly of
     Dirac operator: a holographic derivation}, ArXiv:1111.1463.}
  \ref{CandA}{Cappelli,A. and D'Appollonio, \pl{487B}{2000}{87}.}
  \ref{CandT2}{Copeland,E. and Toms,D.J. \cqg {3}{1986}{431}.}
   \ref{Allais}{Allais, A. {\it JHEP} {\bf 1011} (2010) 040.}
     \ref{Tseytlin}{Tseytlin,A.A. {\it On Partition function and Weyl anomaly of
     conformal higher spin fields} ArXiv:1309.0785.}
     \ref{KPS2}{Klebanov,I.R., Pufu,S.S. and Safdi,B.R. {\it JHEP} {\bf 1110} (2011) 038.}
    \ref{CaandWe}{Candelas,P. and Weinberg,S. \np{237}{1984}{397}.}
     \ref{ChandD}{Chang,P. and Dowker,J.S. \np{395}{1993}{407}.}
 \ref{Steffensen}{Steffensen,J.F. {\it Interpolation}, (Williams and Wilkins,
    Baltimore, 1927).}
     \ref{Barnesa}{Barnes,E.W. {\it Trans. Camb. Phil. Soc.} {\bf 19} (1903) 374.}
    \ref{DowGJMS}{Dowker,J.S.  \jpa{44}{2011}{115402}.}
    \ref{Dowren}{Dowker,J.S. \jpamt {46}{2013}{2254}.}
 \ref{MandD}{Mansour,T. and Dowker,J.S. {\it Evaluation of spherical GJMS determinants},
 2014, Submitted for publication.}
 \ref{GandK}{Gubser,S.S and Klebanov,I.R. \np{656}{2003}{23}.}
     \ref{Dow30}{Dowker,J.S. \prD{28}{1983}{3013}.}
     \ref{Dowcmp}{Dowker,J.S. \cmp{162}{1994}{633}.}
     \ref{DowGJMSE}{Dowker,J.S. {\it Numerical evaluation of spherical GJMS operators
     for even dimensions} ArXiv:1310.0759.}

\end{putreferences}

\bye